\documentclass[
 superscriptaddress,
 amsmath,
 amssymb,
 aps,
 prl,
 floatfix,
 twocolumn,
 longbibliography
]{revtex4-2}

\newcommand{\bra}[1]{\langle #1\rvert}
\newcommand{\ket}[1]{\lvert #1\rangle}

\newcommand{\op}[2]{\ket{#1} \bra{#2}}

\newcommand{\rpd}[1]{\partial_t #1}

\usepackage{graphicx}% Include figure files
\usepackage{dcolumn}% Align table columns on decimal point
\usepackage{bm}% bold math
\usepackage{hyperref}% add hypertext capabilities
\usepackage{tikz}
\usetikzlibrary{calc}

\usepackage{orcidlink}

\usepackage{amsmath}
\usepackage{upgreek}
\usepackage{physics}
\usepackage{isomath}
\usepackage{placeins}
\usepackage{comment}
\usepackage{mathrsfs}

\begin{document}

\title{Fully Collective Superradiant Lasing with Vanishing Sensitivity to Cavity Length Vibrations}
\author{Jarrod T. Reilly\orcidlink{0000-0001-5410-089X}}
\affiliation{JILA, NIST, and Department of Physics, University of Colorado, 440 UCB, Boulder, CO 80309, USA}
\author{Simon B. J\"ager\orcidlink{0000-0002-2585-5246}}
\affiliation{Physikalisches Institut, University of Bonn, Nussallee 12, 53115 Bonn, Germany}
\author{John Cooper}
\affiliation{JILA, NIST, and Department of Physics, University of Colorado, 440 UCB, Boulder, CO 80309, USA}
\author{Murray J. Holland\orcidlink{0000-0002-3778-1352}}
\affiliation{JILA, NIST, and Department of Physics, University of Colorado, 440 UCB, Boulder, CO 80309, USA}

\date{\today}


\begin{abstract}
To date, realization of a continuous-wave active atomic clock has been elusive primarily due to parasitic heating from spontaneous emission while repumping the atoms. Here, we propose a solution to this problem by replacing the random emission with coupling to an auxiliary cavity, making repumping a fully collective process. While it is known that collective two-level models do not possess a generic lasing threshold, we show this restriction is overcome with multi-level atoms since collective pumping and decay can be performed on distinct transitions. Using relevant atomic parameters, we find this system is capable of producing an $\mathcal{O}$(100 $\mu$Hz)-linewidth continuous-wave superradiant laser. Our principal result is the potential for an operating regime with cavity length vibration sensitivity below $\mathcal{O}(10^{-14} / g)$, including a locus of parameter values where it completely vanishes even at steady-state.
\end{abstract}

{
\let\clearpage\relax
\maketitle
}

\emph{Introduction.---}
Throughout the history of physics, improvements in precision measurement technologies have enabled advances in  understanding the basic building blocks of the universe~\cite{Aasi,Abbott,Abbott,Abi,Albahri,Akiyama,Abud,Bothwell,Agazie,Roussy,Ye}. 
Nowadays, one of the most cutting-edge metrological devices is the optical lattice atomic clock~\cite{Derevianko,Ludlow}. 
This device leverages metastable states of atoms to produce extreme accuracy and sensitivity for precise time keeping and metrology~\cite{Derevianko,Ludlow,Hinkley,PedrozoPenafiel,Bothwell,Robinson,Yang}. 
With the increasing accuracy of atomic clocks, a new frontier emerges in which experiments will be able to probe spacetime curvature over the size of the finite wave function of a quantum object~\cite{Bothwell}. However, to push back the frontier further it will be necessary to overcome the constraints placed by the stability of the laser with which the atomic oscillator is read out. The current optimal engineering solution to this problem is the result of decades of development of ultrastable reference cavities whose scope is confined to specialized labs. This requirement limits the widespread application of the most precise atomic clocks and prohibits deployment in adverse environments.

Recently, there has been growing interest in laboratories around the world in an alternate solution aimed at developing \emph{active} atomic clocks~\cite{Chen,Meiser,Meiser2,Meiser3,Bohnet,Maier,Kraemer,Norcia,Norcia2,Norcia3,Norcia4,Debnath,Zhang,Laske,Schaeffer,Zhang,Pan,Liu,Jager3,Tang,Kristensen,PoncianoOjeda,Fama}. 
Such active clocks are synonymous with superradiant lasers~\cite{Meiser}, and do not need a reference laser as they intrinsically generate their own phase-stable light~\cite{Norcia2}.
Since the phase coherence of the output field is stored in the atoms rather than the cavity photons, superradiant lasers can become incredibly robust to fluctuations of the cavity length that arise from thermal or acoustic noise sources~\cite{Bohnet,Norcia3}. Such noise sources are the principal deleterious factors that limit the stability of state-of-the-art reference lasers~\cite{Ludlow2,Hall,Kessler,Zhang5,Robinson2,Kedar}. 
Despite their promise, active-clock models have been difficult to realize experimentally outside the pulsed regime~\cite{Norcia2,Schaeffer,Kristensen,Fama}.
A primary obstacle to sustaining the superradiant lasing process is atomic heating arising from the random momentum kicks due to spontaneous emission during repumping~\cite{Norcia}. 
It should be emphasized that it is not possible to simply overcome the heating by employing cavity-assisted transitions in these systems. 
This is because any fully collective repumping and fully collective decay on the same transition only generates rotations of the effective two-level system (more explicitly, is restricted to $\mathrm{SU} (2)$ group actions), and these rotations do not change the length of the collective dipole, resulting in the absence of a generic steady-state lasing threshold~\cite{Haake}. 

In this Letter, we propose a method that uses multi-level effects to allow the fixed-length constraint on the collective dipole to be overcome. 
By adding one additional ground state to the system, collective pumping and collective decay can be performed on different transitions leading to SU(3) group actions [instead of SU(2)]. 
We find that it is then possible to achieve continuous-wave superradiant lasing over a large realizable parameter space with a cavity-assisted repumping process.
The lasing light displays the ultracoherence property of former superradiant lasing models~\cite{Meiser,Liu}, with a linewidth that is determined by the cavity-modified spontaneous emission rate from the ultranarrow transition, $\Gamma_c$, which can be sub-mHz~\cite{Norcia2,Muniz}. 
Moreover, we find vanishing sensitivity to variations of the cavity length, i.e., vanishing cavity pulling, which is not possible in previous superradiant laser models at steady-state~\cite{Norcia3,Liu}. 
Lastly, we demonstrate that the fully collective system can survive the adverse effects of single-particle spontaneous emission, resulting in an $\mathcal{O}(100 \, \mu \mathrm{Hz})$-linewidth continuous-wave laser on an $\mathcal{O}(\mathrm{mHz})$ transition. 

\begin{figure}[!t]
    \centerline{\includegraphics[width=\linewidth]{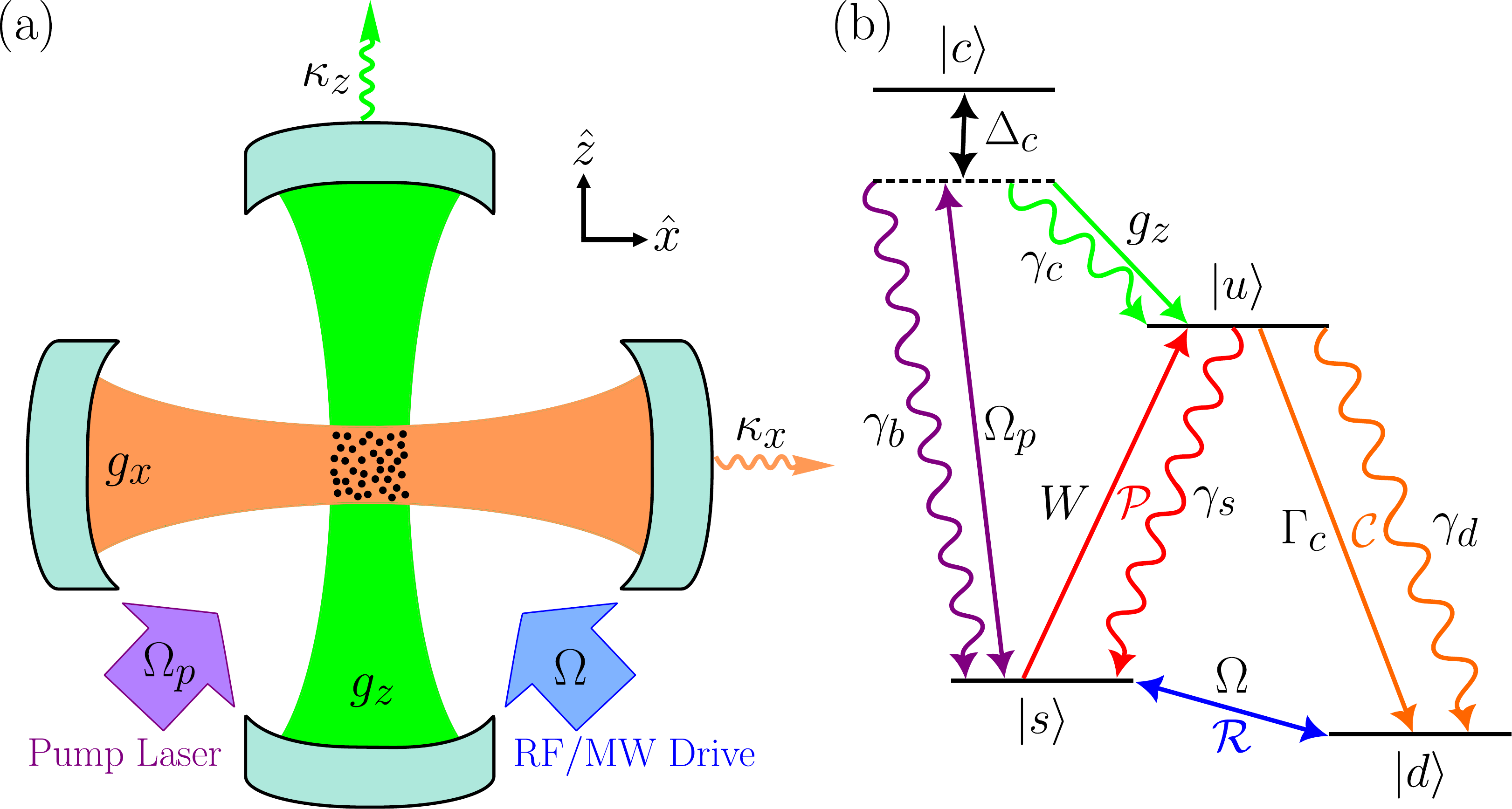}}
    \caption{The SU(3) superradiant laser. 
    (a) Experimental configuration based on a crossed-trap geometry of two leaky optical resonators with two external coherent drives. 
    (b) The four-level atomic configuration, symbols explained in text.}
    \label{Schematic}
\end{figure}

\emph{Model.---}
Figure~\ref{Schematic}(a) shows a crossed-trap geometry of the fully collective SU(3) superradiant laser.
We consider a system of $N$ alkaline-earth(-like) atoms that couple to two lossy optical cavities, or alternatively to two modes of a single cavity.
As depicted in Fig.~\ref{Schematic}(b), the atoms have an internal $\Lambda$-configuration with electronic ground states $\ket{d}$ and $\ket{s}$ and excited state $\ket{u}$, as well as one additional auxiliary excited state~$\ket{c}$ to mediate pumping.
%The annotation here is motivated by the four lowest energy quarks that share a related group structure.
The $\ket{d} \leftrightarrow \ket{u}$ transition is the ultranarrow clock transition.
The cavity modes, $\hat{a}_x$ and $\hat{a}_z$, exchange photons with the atoms with coupling constants $g_x$ and~$g_z$. 
The $x$-cavity is resonant with the $\ket{d} \leftrightarrow \ket{u}$ transition, while the $z$-cavity drives $\ket{c} \leftrightarrow \ket{u}$ off-resonantly with a detuning~$\Delta_c$. 
The atoms are driven by an optical field $\Omega_p$ that couples $\ket{s} \leftrightarrow \ket{c}$, also with detuning~$\Delta_c$, and a radio frequency or microwave field~$\Omega$ that couples the two ground states $\ket{d} \leftrightarrow \ket{s}$.

The dissipative effects we consider stem from the system’s coupling to free-space electromagnetic modes. 
We include spontaneous emission from $\ket{c} \rightarrow \ket{u}$ and $\ket{c} \rightarrow \ket{s}$ at rates $\gamma_c$ and $\gamma_b$, as well as $\ket{u} \rightarrow \ket{d}$ and $\ket{u} \rightarrow \ket{s}$ at rates $\gamma_d$ and $\gamma_s$. 
The cavity mode decay rates, $\kappa_x$ and $\kappa_z$, describe the output coupling of the photons. We do not include any optomechanical effects here as we assume the atoms are tightly confined in a supporting optical lattice for each cavity direction (not shown). 
In this case, the system is symmetric to permutations of atom indices.

We assume a separation of time scales in our parameters such that we can adiabatically eliminate parts of the system to find effective equations of motion of an essential model. 
In particular, we assume the detuning of $\hat{a}_z$ and $\Omega_p$ from $\ket{c}$ is large, $\abs{\Delta_c} \gg \Omega_p, \sqrt{N} g_z$ such that we can eliminate the auxiliary state $\ket{c}$~\cite{Reiter}. 
The adiabatic elimination of $\ket{c}$ gives rise to an effective resonant two-photon Raman transition consisting of $\Omega_p$ and the z-cavity. 
We then assume the bad-cavity limit, $\kappa_x \gg \sqrt{N} g_x$ and $\kappa_z \gg \sqrt{N} \abs{\Delta_c} g_z \Omega_p / [2 \Delta_c^2 + \left( \gamma_c + \gamma_b \right)^2 / 2]$, such that we can eliminate both cavity modes~\cite{Jager}. 
This results in a simplified master equation for the reduced atomic density matrix $\rpd \hat{\rho} = \hat{\mathcal{L}} \hat{\rho}$ with the Liouvillian,
\begin{equation} \label{Liouvillian}
    \hat{\mathcal{L}} \hat{\rho} = i \Omega \left[ \hat{\rho}, \hat{\mathcal{R}}_x \right] + \hat{\mathcal{D}} \left[ \sqrt{\Gamma_c} \, \hat{\mathcal{C}}_- \right] \hat{\rho} + \hat{\mathcal{D}} \left[ \sqrt{W} \hat{\mathcal{P}}_+ \right] \hat{\rho} + \hat{\mathcal{L}}_{\mathrm{sp}} \hat{\rho},
\end{equation} 
where $\hat{\mathcal{D}} [\hat{O}] \hat{\rho} = \hat{O} \hat{\rho} \hat{O}^{\dagger} - [\hat{O}^{\dagger} \hat{O} \hat{\rho} + \hat{\rho} \hat{O}^{\dagger} \hat{O}] / 2$ is the Lindblad superoperator with the effective pumping and decay rates $W$ and $\Gamma_c$.
Here, we have defined the collective raising operators $\hat{\mathcal{C}}_+ = \sum_j \hat{\sigma}_{ud}^{(j)}$, $\hat{\mathcal{P}}_+ = \sum_j \hat{\sigma}_{us}^{(j)}$, and $\hat{\mathcal{R}}_+ = \sum_j \hat{\sigma}_{ds}^{(j)}$, where $\hat{\sigma}_{ab}^{(j)} \equiv \op{a}{b}_j$ is a Pauli operator for atom $j$, as well as $\hat{\mathcal{R}}_x = (\hat{\mathcal{R}}_+ + \hat{\mathcal{R}}_-) / 2$. Moreover, the single-particle decoherence Liouvillian,
\begin{equation}
    \begin{aligned}
\hat{\mathcal{L}}_{\mathrm{sp}} = \sum_{j = 1}^N \Big( &\hat{\mathcal{D}} \left[ \sqrt{\gamma_d} \hat{\sigma}_{du}^{(j)} \right] + \hat{\mathcal{D}} \left[ \sqrt{\gamma_s} \hat{\sigma}_{su}^{(j)} \right] \\
+ &\hat{\mathcal{D}} \left[ \sqrt{w} \hat{\sigma}_{us}^{(j)} \right] + \hat{\mathcal{D}} \left[ \sqrt{\gamma_p} \hat{\sigma}_{ss}^{(j)} \right] \Big),
    \end{aligned}
\end{equation}
models spontaneous emission from $\ket{u}$, as well as single-particle pumping $w$ and dephasing $\gamma_p$ stemming from incoherent Raman and Rayleigh scattering off $\ket{c}$, respectively~\cite{Shankar}. 
Further details on the derivation of the model are presented in the Supplemental Material (SM)~\cite{suppMat}.

For clarity, we begin by analyzing the regime in which the collective processes, which scale with the atom number, dominate over all single-particle processes, so that $\hat{\mathcal{L}}_{\mathrm{sp}}$ can be considered small and be neglected. 
In this case, the model consists of three steps. 
First, atoms in $\ket{s}$ are collectively pumped to $\ket{u}$ via the $z$-cavity-assisted Raman process with rate $N W$. Then, the atoms in $\ket{u}$ collectively decay via the $x$-cavity at a rate $N \Gamma_c$. 
The $\Omega$ drive completes the cycle by coupling the ground states so atoms can return to $\ket{s}$. 

The permutation symmetry of the atoms gives the collective parts of the master equation an SU(3) group structure, and this allows the default exponential scaling of the Liouville state space, $3^{2 N}$, to be reduced to a polynomial scaling, $N^4$~\cite{Reilly3}. 
This enables exact simulations of the master equation to be performed for a substantial number of atoms $N \gg 1$ without resorting to approximation methods. 
By noting the master equation has an additional $\mathrm{U} (1)$ symmetry under the arbitrary rotation $\hat{\rho} \rightarrow \exp[i \theta \sum_j \hat{\sigma}_{uu}^{(j)}] \hat{\rho} \exp[- i \theta \sum_j \hat{\sigma}_{uu}^{(j)}]$, the scaling can be reduced further to $N^3$ allowing efficient access to  relevant one-time observables, the laser linewidth, and photon bunching statistics~\cite{Reilly3,suppMat}. 
By developing this fully quantum simulation, we can find the steady-state (ss) density matrix by calculating the kernel of the Liouvillian, $\hat{\mathcal{L}} \hat{\rho}_{\mathrm{ss}} = 0$. 

\emph{Signatures of superradiant lasing.---}
The hallmark of superradiance is the constructive interference of the photon emission amplitudes from separate emitters that results in the formation of a macroscopic dipole~\cite{Gross}. 
In bad-cavity superradiant models, it occurs in concert with spin synchronization which allows a $N^2$ scaling of the output power since the cavity photon amplitude is effectively phase locked to the collective dipole oscillation. 
For the SU(3) superradiant lasing model, the primary interest is the output of the $x$-cavity, and due to $\hat{a}_x \sim \hat{\mathcal{C}}_-$ after adiabatic elimination~\cite{suppMat}, the signature of steady-state superradiance on the $\ket{d} \leftrightarrow \ket{u}$ clock transition is given by $\bigl<\hat{\mathcal{C}}_+ \hat{\mathcal{C}}_-\bigr>_{\mathrm{ss}} \propto N^2$.
Furthermore, one can examine whether the system is lasing by calculating the second-order coherence function at zero time delay, $g^{(2)} (0) = \bigl<\hat{\mathcal{C}}_+ \hat{\mathcal{C}}_+ \hat{\mathcal{C}}_- \hat{\mathcal{C}}_-\bigr>_{\mathrm{ss}} / \bigl<\hat{\mathcal{C}}_+ \hat{\mathcal{C}}_-\bigr>_{\mathrm{ss}}^2$, which becomes unity for coherent light~\cite{Mandel}. 

\begin{figure}[h]
    \centerline{\includegraphics[width=\linewidth]{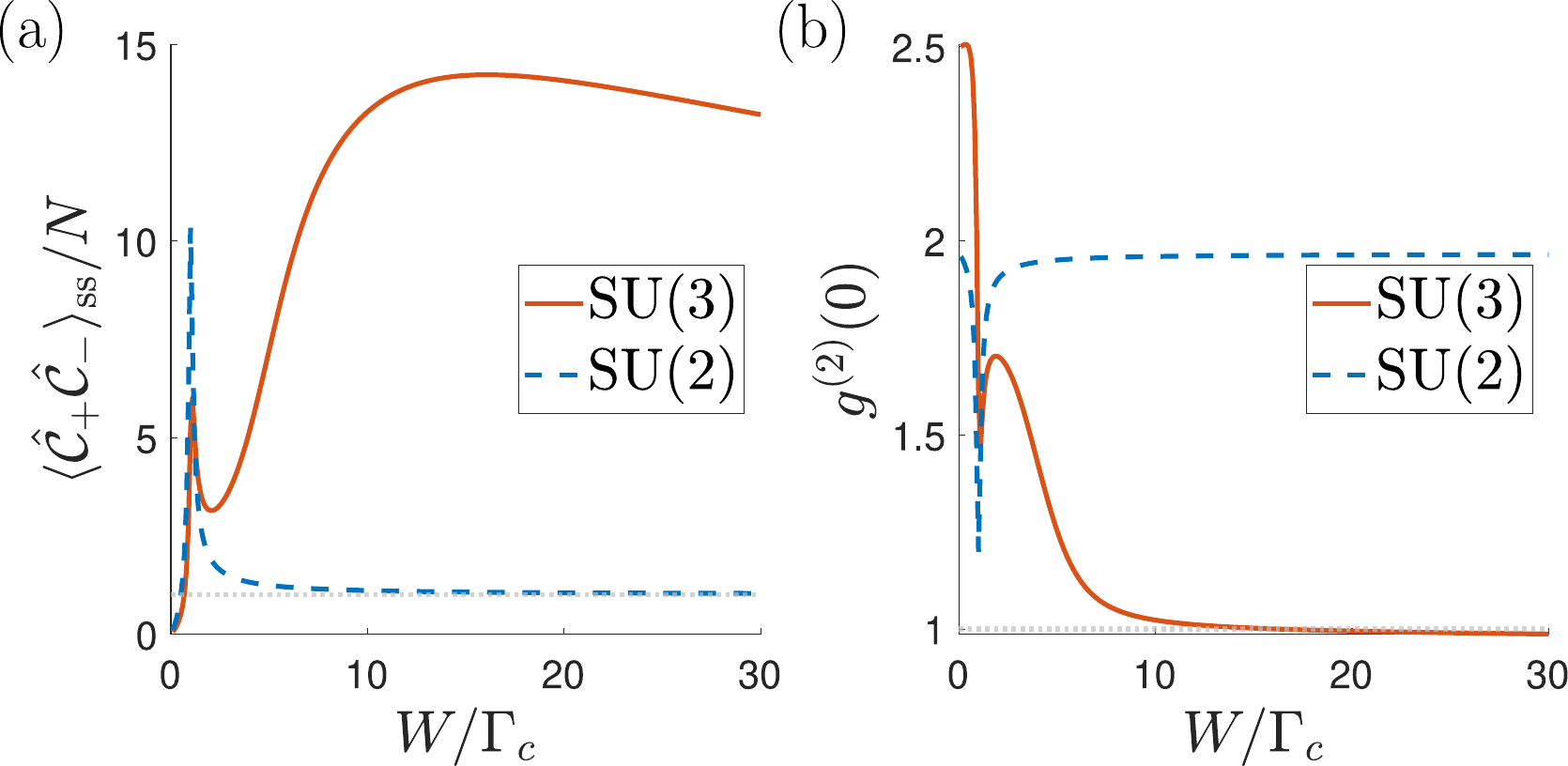}}
    \caption{Comparison between the fully collective $\mathrm{SU} (2)$ (from~\cite{Haake}) and $\mathrm{SU} (3)$ (this paper) superradiant lasing models at steady-state. Here, $N = 60$, $\Omega = 1.9 N \Gamma_c \approx \sqrt{15} N \Gamma_c / 2$, and $\hat{\mathcal{L}}_{\mathrm{sp}} = 0$. (a) Behavior of the photon output flux showing a broad regime of lasing in the multi-level system. (b) The second-order coherence, which is unity for coherent light. }
    \label{SU2vSU3Fig}
\end{figure}

Figure~\ref{SU2vSU3Fig} displays the steady-state intensity $\bigl<\hat{\mathcal{C}}_+ \hat{\mathcal{C}}_-\bigr>_{\mathrm{ss}}$ and $g^{(2)} (0)$ as a function of pump rate $W$. 
To demonstrate why the multiple ground state atomic system is necessary, we compare our results with the fully collective $\mathrm{SU} (2)$ model~\cite{Haake,Haake2} given by the master equation $\rpd \hat{\rho}_2 = \hat{\mathcal{D}} \left[ \sqrt{\Gamma_c} \, \hat{\mathcal{C}}_- \right] \hat{\rho}_2 + \hat{\mathcal{D}} \left[ \sqrt{W} \hat{\mathcal{C}}_+ \right] \hat{\rho}_2$ with no $\ket{s}$ state. 
While both models exhibit an intensity spike at $W = \Gamma_c$, the SU(2) model only obtains steady-state superradiance near this one point. 
In contrast, the SU(3) model exhibits a whole plateau of steady-state superradiance when $W \gg \Gamma_c$.
Similarly, we see that even at the superradiant spike at $W \approx \Gamma_c$, the atom number $N = 60$ is not large enough for the $\mathrm{SU} (2)$ model to exhibit coherent light. 
In the thermodynamic limit $N \rightarrow \infty$, the SU(2) model does achieve $g^{(2)} (0) = 1$ but only at the singular point $W = \Gamma_c$, and thus we refer to the SU(2) system as lacking a generic lasing transition. 
This is not the case for the SU(3) system, where $g^{(2)} (0) \approx 1$ throughout the steady-state superradiance plateau. 

\begin{figure*}
    \centerline{\includegraphics[width=\linewidth]{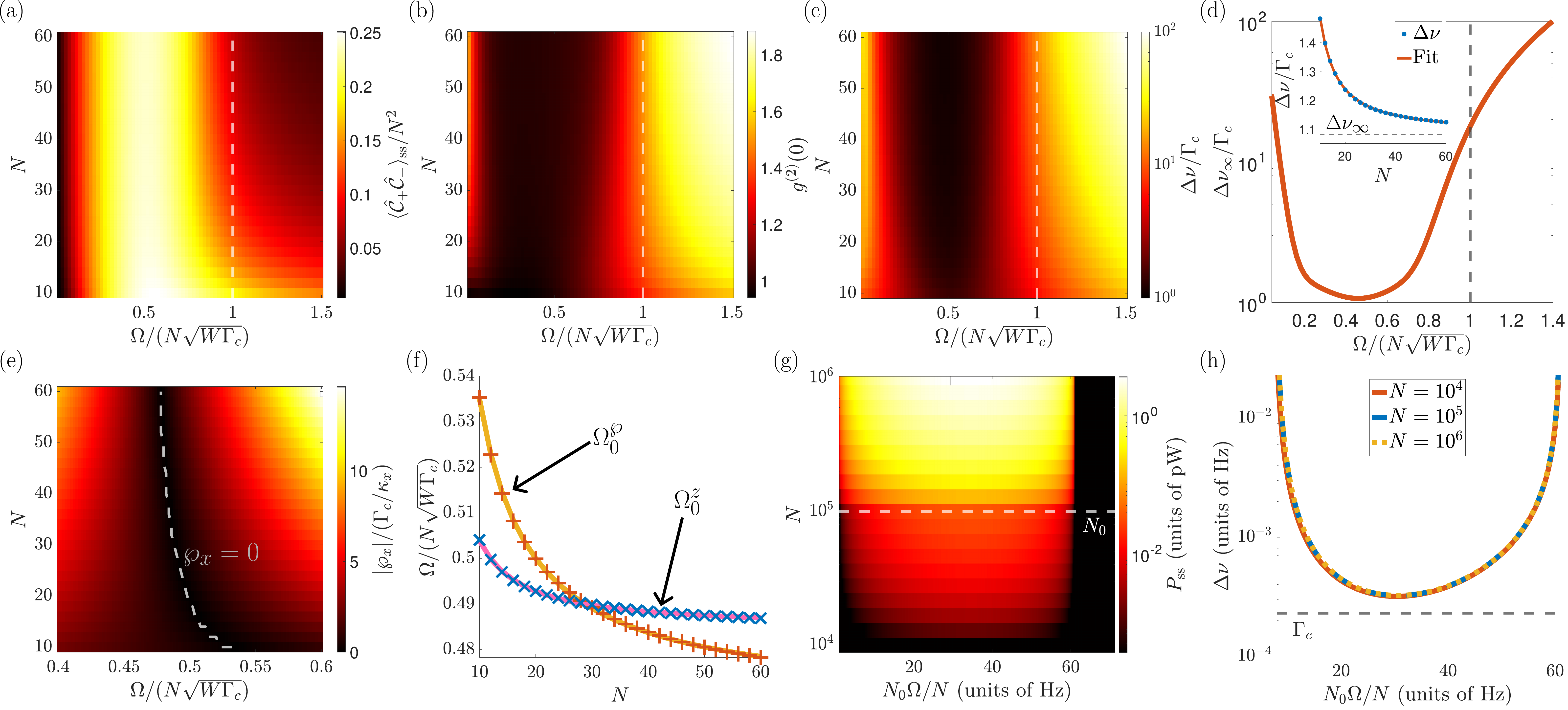}}
    \caption{Steady-state properties of the $\mathrm{SU}(3)$ superradiant laser at $W = 15 \Gamma_c$. 
    In (a)-(f), we use $\hat{\mathcal{L}}_{\mathrm{sp}} = 0$, while (g)-(h) use the $\hat{\mathcal{L}}_{\mathrm{sp}}$ rates for Ba outlined in the SM~\cite{suppMat} as well as a reference atom number $N_0 = 10^5$ to keep consistent x-axes. 
    (a) Scaled output intensity. 
    (b) Photon bunching statistics. 
    (c)-(d) Laser linewidth. 
    (e) Cavity pulling coefficient and locus of vanishing cavity length sensitivity (dashed gray line). 
    (f) Pump intensity for zero pulling coefficient ($\Omega_0^{\wp}$) and zero $\ket{d} \leftrightarrow \ket{u}$ inversion ($\Omega_0^z$). 
    (g) Output power for Ba parameters. 
    (h) Laser linewidth for Ba parameters. 
    Note that in (f) and the inset of~(d), we fit the respective data as a function of $N$ to $f(N) = X + Y / N + Z / N^2$ to extract the thermodynamic limit $X$.
    The inset in (d) shows the fit for the point $\Omega = N \sqrt{W \Gamma_c} / 2$.}
    \label{FullLasingFig}
\end{figure*}

Using a mean-field analysis, we can extract the threshold where the $\mathrm{SU} (3)$ model predicts the steady-state lasing phase transition. 
We find that there are three different parameter regimes, characterized by inequalities of $\Omega$ compared to $N \Gamma_c$ and $N \Gamma_c / 2$, which lead to different superradiant lasing thresholds (see SM~\cite{suppMat}).
However, if we focus on the case $W > \Gamma_c$ as considered below, then lasing occurs when $\Omega < N \sqrt{W \Gamma_c}$. 
We explore this lasing transition in Figs.~\ref{FullLasingFig}(a)-(d) for the case $W = 15 \Gamma_c$. 
In Figs.~\ref{FullLasingFig}(a) and~(b), we again examine the steady-state intensity and $g^{(2)} (0)$ where there is a clear superradiance transition around $\Omega = N \sqrt{W \Gamma_c}$. 
Here, the intensity goes from a $N$ scaling on the right to a $N^2$ scaling on the left, reaching a maximum value of $N^2 / 4$. 
Similarly, this presents the characteristics of a true lasing transition as the second-order coherence function goes from bunched light [$g^{(2)} (0) > 1$] on the right to Poisson statistics [$g^{(2)} (0) = 1$] on the left.

To explore the utility of the system for the application as an active clock, we evaluate the linewidth $\Delta \nu$ of the output lasing light. 
This requires determination of the first-order coherence function $g^{(1)} (\tau) \propto \bigl<\hat{\mathcal{C}}_+ (\tau) \hat{\mathcal{C}}_- (0)\bigr>_{\rm ss}$~\cite{Mandel}, whose Fourier transform gives the spectrum $S(\tilde{\omega}) = 2 \Re \left[ \int_0^{\infty} d \tau e^{i \tilde{\omega} \tau} g^{(1)} (\tau) \right]$.
Since we anticipate a Lorentzian-shaped spectrum, we extract the laser's frequency $\omega$ and linewidth $\Delta \nu$ by fitting the parametrized function $g^{(1)} (\tau) \propto \exp[(i \omega - i \tilde{\omega}_d - \Delta \nu / 2) \tau]$. 
The transition frequency of the gain medium $\tilde{\omega}_d$ appears since we are in a rotating frame~\cite{suppMat}. 
The fit parameters can be deduced from the first non-zero eigenvalue $\lambda_1$ of $\hat{\mathcal{L}}$ since $\omega = \Im[\lambda_1] + \tilde{\omega}_d$ and $\Delta \nu = - 2 \Re[\lambda_1]$. 
In Fig.~\ref{FullLasingFig}(c), we display how $\Delta \nu$ varies over the $\Omega = N \sqrt{W \Gamma_c}$ phase transition. 
Here, we see a rapid contraction of the linewidth in the lasing regime as the linewidth goes from $\mathcal{O}(10^2 \Gamma_c)$ on the right to $\mathcal{O}(\Gamma_c)$ on the left. 
This is because of a cancellation that occurs between two extensive terms in the analytic theory of the linewidth once the atoms synchronize into a macroscopic dipole, similar to previous superradiant lasing models~\cite{Meiser,Liu}. 
To explore the large $N$ behavior of the linewidth, in Fig.~\ref{FullLasingFig}(d) we fit the linewidth as a function of $N$ to $f(N) = \Delta \nu_{\infty} + B / N + C / N^2$ and take $N \rightarrow \infty$ such that $\Delta \nu \approx \Delta \nu_{\infty}$. 
The inset demonstrates that this fit aligns well with the simulation data. 
We see a change of the scaling of $\Delta \nu_{\infty}$ once we cross the phase transition point.
Furthermore, we see that around $\Omega \approx N \sqrt{W \Gamma_c} / 2$, $\Delta \nu_{\infty}$ approaches the decay rate $\Gamma_c$. 

\emph{Cavity pulling.---}
We now focus on the dependence of the frequency of the coherent light to changes of the $x$-cavity length. 
To do this, we define the detuning of the $x$-cavity from the gain medium's transition frequency, $\Delta_x = \tilde{\omega}_d - \omega_x$. 
This adds a Hamiltonian term, $i \chi_x \left[ \hat{\mathcal{C}}_+ \hat{\mathcal{C}}_-, \hat{\rho} \right]$ with $\chi_x \approx \Delta_x \Gamma_c / \kappa_x$, to Eq.~\eqref{Liouvillian}~\cite{suppMat}. 
We can then fit the laser's frequency $\omega$ as a linear function of $\chi_x$ to find the cavity pulling $\wp_x = (\tilde{\omega}_d - \omega) / \Delta_x$. 
In Fig.~\ref{FullLasingFig}(e), we show that there is a broad region of $\Omega$ near the minimum linewidth where the cavity pulling becomes $\abs{\wp_x} = \mathcal{O}(10 \Gamma_c / \kappa_x)$ which is in general quite small in the bad-cavity regime. 
For example, Ref.~\cite{Norcia2} reports $\Gamma_c = \mathcal{O}(100 \, \mu \mathrm{Hz})$ and $\kappa_x = \mathcal{O}(100 \, \mathrm{kHz})$ such that $\abs{\wp_x} =  \mathcal{O}(10^{-8})$, which shows the potential to achieve a pulling coefficient that is a million times lower than the current steady-state record of $0.0148$~\cite{Zhang4}. 
We also find that a point exists for each $N$ where $\wp_x = 0$ (gray line).
To see if this behavior persists in the thermodynamic limit $N \rightarrow \infty$, we find the $\Omega$ where $\wp_x = 0$, which we label $\Omega_0^{\wp}$, as a function of $N$ and again fit this to the function $f(N) = X + Y / N + Z / N^2$. 
We display the simulation data and the fit (yellow line) in Fig.~\ref{FullLasingFig}(f) where we find it asymptotes to $\Omega_0^{\wp} \approx 0.469 N \sqrt{W \Gamma_c}$. 

It should be emphasized that the small or vanishing pulling coefficient values are possible in this system due to its multi-level structure which allows the system to lase without inversion on the $\mathcal{C}$ transition~\cite{PhysRev.107.1579}. This is because the lasing threshold only relies on inversion between dressed states, i.e., the eigenstates of the nondissipative dynamics of the full system. 
Looking at the Heisenberg-Langevin equation for the dipole $\rpd \hat{\mathcal{C}}_- = - 2 i \chi_x \hat{\mathcal{C}}_z \hat{\mathcal{C}}_- + \ldots$, where $\hat{\mathcal{C}}_z = [ \hat{\mathcal{C}}_+, \hat{\mathcal{C}}_- ] / 2$, we see that this insensitivity to $\Delta_x$ is a result of $\bigl<\hat{\mathcal{C}}_z\bigr>_{\mathrm{ss}}$ becoming small when going from positive to negative as a function of $\Omega$ (see SM~\cite{suppMat}). 
To this end, we display the $\Omega$ value where $\bigl<\hat{\mathcal{C}}_z\bigr>_{\mathrm{ss}} = 0$, labeled $\Omega_0^z$, in Fig.~\ref{FullLasingFig}(f).
The $f(N)$ fit (pink line) gives $\Omega_0^z \approx 0.485 N \sqrt{W \Gamma_c}$ in the thermodynamic limit. 
Unlike the pulsed superradiance schemes of Refs.~\cite{Norcia3,Norcia4}, we find $\Omega_0^{\wp} \neq \Omega_0^z$ due to additional frequency shifts from couplings with the $\mathcal{P}$ and $\mathcal{R}$ dipoles. 

\emph{Single-particle decoherence.---} Single-particle decoherence can remove the state from the symmetric subspace of $\mathrm{SU} (3)$. 
We now investigate whether the fully collective superradiant lasing mechanism can survive the adverse effect of $\hat{\mathcal{L}}_{\mathrm{sp}}$ which can lower the eigenvalue of the cubic Casimir operator, and thus limit the maximum achievable dipole length on the lasing transition. 
Here, Casimir operator refers to an invariant of the Lie algebra, which can be thought of as a generalization of the total angular momentum operator from SU(2).
To find the steady-state of the Liouvillian Eq.~\eqref{Liouvillian} with $\hat{\mathcal{L}}_{\mathrm{sp}} \neq 0$, we now assume $\hat{\rho}$ is factorizable such that we can employ a mean-field approximation including Langevin noise~\cite{Gardiner}. 
The details of this approximation are shown in the SM~\cite{suppMat}. 
This approximation method also allows us to simulate relevant atom numbers, e.g., $N = \mathcal{O}(10^5)$ as in Refs.~\cite{Norcia2,Norcia3}. 

For experimental context, we consider the $1085 \, \mathrm{nm}$ ${}^1 S_0 \leftrightarrow {}^3 D_2$ clock transition of Ba~\cite{Dzuba2}, along with the auxiliary state ${}^3 P_1$. 
The odd isotopes of Ba have non-zero nuclear spin, and so hyperfine structure in ${}^1 S_0$ for $\ket{d}$ and $\ket{s}$. 
We use cavity cooperativity parameters of $\mathscr{C}_x = \Gamma_c / \gamma_d = 0.1$ and $\mathscr{C}_z = W / w = 0.1$ which are obtainable~\cite{Norcia2}.
The linewidth of the clock transition is~\cite{Dzuba2} $\gamma_d = \gamma_s = 10 \Gamma_c = 2.3 \, \mathrm{mHz}$, and choosing $W = 15 \Gamma_c$ leads to $w = 10 W = 34.5 \, \mathrm{mHz}$.
Finally, the branching ratios from ${}^3 P_1$ gives~\cite{Dzuba2} $\gamma_p \approx 1.2 w = 41.5 \, \mathrm{mHz}$, which is a particularly advantageous property.
Note that the $\mathrm{SU} (2)$ model from Ref.~\cite{Meiser} operates with $w_2 = \mathcal{O}(N \Gamma_c / 2)$, and so our $\mathrm{SU} (3)$ model reduces the effective scattering rate off $\ket{c}$ by a factor $\mathcal{O}(N \mathscr{C}_z \Gamma_c / [2 W]) \sim N / 100$. 
We therefore expect our system to be subject to reduced detrimental heating when compared with prior models.

Using these parameters, we display the steady-state power $P_{\mathrm{ss}} = \hbar \omega \Gamma_c \bigl<\hat{\mathcal{C}}_+ \hat{\mathcal{C}}_-\bigr>_{\mathrm{ss}}$ and linewidth in Figs.~\ref{FullLasingFig}(g) and~3(h).
We again assume $\Delta_x = 0$ such that $\omega = \tilde{\omega}_d$. 
The same phase transition from superradiant lasing to non-lasing occurs at large $\Omega$, although the critical point is shifted from $\Omega = N \sqrt{W \Gamma_c}$ to $\Omega \approx 0.675 N \sqrt{W \Gamma_c}$ due to $\hat{\mathcal{L}}_{\mathrm{sp}}$. 
The maximum intensity of $\bigl<\hat{\mathcal{C}}_+ \hat{\mathcal{C}}_-\bigr>_{\mathrm{ss}} \approx 0.09 N^2$ occurs at $\Omega \approx 0.39 N \sqrt{W \Gamma_c}$, and so the effects of $\hat{\mathcal{L}}_{\mathrm{sp}}$ have only decreased the maximum intensity by a factor of $3$ for the chosen parameters.  
For $N = 10^6$, we find a maximum power of $P_{\mathrm{ss}} \approx 3.75 \, \mathrm{pW}$ which should be a sufficient photon flux to achieve a servo lock as an optical phase reference~\cite{Meiser}. 
Meanwhile, the linewidth is minimized around $\Omega \approx 0.35 N \sqrt{W \Gamma_c}$ where it reaches $\Delta \nu \approx 1.4 \Gamma_c = 325 \, \mu \mathrm{Hz}$, which stays consistent for large~$N$. If realized, this would have higher stability than any laser demonstrated to date~\cite{Matei}.  
We find there is a $4$ Hz range, $258 \, \mathrm{Hz} \lesssim \Omega \lesssim 262 \, \mathrm{Hz}$, where the pulling coefficient becomes $\abs{\wp_x} = \mathcal{O}(10^{-6})$ when $N = 10^6$ (see SM~\cite{suppMat}), where we have again used $\kappa_x = \mathcal{O}(100 \, \mathrm{kHz})$~\cite{Norcia2}. 
If we assume that the $x$-cavity is a Fabry-P\'erot, with vibrational sensitivity of $K = \mathcal{O}(10^{-8} / \mathrm{g})$~\cite{Chen2}, the resulting effective sensitivity over this entire range would be~\cite{Liu} $\abs{\wp_x} K = \mathcal{O}(10^{-14} / \mathrm{g})$. 
This would outperform state-of-the-art cavity-stabilized lasers where $K_{\mathrm{st}} = \mathcal{O}(10^{-12} / \mathrm{g})$~\cite{Robinson2,Kedar}.
Lastly, we find $\Omega^{\wp}_0 \approx 260 \, \mathrm{Hz}$ when $N = 10^6$~\cite{suppMat}, showing that $\wp_x$ still vanishes even in the presence of $\hat{\mathcal{L}}_{\mathrm{sp}}$.

In summary, the SU(3) superradiant laser brings additional capabilities that are not present in earlier models that do not possess multiple ground states. We find two potential advantages; the possibility for collective pumping of an auxiliary cavity mode, and the possibility for reduced or vanishing cavity pulling. If realized, construction of active atomic clocks founded on these ideas could potentially usher in a new paradigm of optical coherence. 

\begin{acknowledgments}
We would like to thank Francesca Fam\`a, John D. Wilson, and Catie LeDesma for useful discussions.
We acknowledge support from NSF OMA 2016244; NSF PHY Grant No. 2207963; and NSF PFC 2317149. 
S. B. J. acknowledges support from the Deutsche Forschungsgemeinschaft (DFG, German Research Foundation) under project number 277625399 - TRR 185 (B4) and under Germany’s Excellence Strategy – Cluster of Excellence Matter and Light for Quantum Computing (ML4Q) EXC 2004/1 – 390534769.
\end{acknowledgments}
\vspace*{-1pc}

\bibliography{references.bib}

\appendix

\end{document}

%\preprint{APS/123-QED}

\title{Supplemental Material: Fully Collective Superradiant Lasing with Vanishing Sensitivity to Cavity Length Vibrations}
\author{Jarrod T. Reilly\orcidlink{0000-0001-5410-089X}}
\affiliation{JILA, NIST, and Department of Physics, University of Colorado, 440 UCB, Boulder, CO 80309, USA}
\author{Simon B. J\"ager\orcidlink{0000-0002-2585-5246}}
\affiliation{Physikalisches Institut, University of Bonn, Nussallee 12, 53115 Bonn, Germany}
\author{John Cooper}
\affiliation{JILA, NIST, and Department of Physics, University of Colorado, 440 UCB, Boulder, CO 80309, USA}
\author{Murray J. Holland\orcidlink{0000-0002-3778-1352}}
\affiliation{JILA, NIST, and Department of Physics, University of Colorado, 440 UCB, Boulder, CO 80309, USA}

\date{\today}


\maketitle

\tableofcontents

\section{Derivation of SU(3) Superradiant Lasing Model}
In this section, we derive the SU(3) superradiant lasing model studied in the main text. 

\subsection{Double cavity system}
Based on the couplings depicted in Fig.~1 of the main text, we begin with the Hamiltonian
\begin{equation}
    \begin{aligned}
\hat{H}_1 = & \hbar \omega_x \hat{a}_x^{\dagger} \hat{a}_x + \hbar \omega_z \hat{a}_z^{\dagger} \hat{a}_z + \sum_{j = 1}^N \Bigg[ \hbar \omega_c \hat{\sigma}_{cc}^{(j)} - \hbar \omega_d \hat{\sigma}_{dd}^{(j)} - \hbar \omega_s \hat{\sigma}_{ss}^{(j)} \\
&+ \hbar g_x \left( \hat{\sigma}_{ud}^{(j)} \hat{a}_x + \mathrm{H.c.} \right) + \hbar g_z \left( \hat{\sigma}_{cu}^{(j)} \hat{a}_z + \mathrm{H.c.} \right) \\
&+ \frac{\hbar \Omega}{2} \left( \hat{\sigma}_{sd}^{(j)} e^{i \omega_r t} + \mathrm{H.c.} \right) + \frac{\hbar \Omega_p}{2} \left( \hat{\sigma}_{sc}^{(j)} e^{i \omega_p t} + \mathrm{H.c.} \right) \Bigg],
    \end{aligned}
\end{equation}
where $\hat{\sigma}_{ab}^{(j)} = \op{a}{b}_j$ is a Pauli operator for atom $j$. 
Here, we have defined the $x$-cavity and $z$-cavity frequencies, $\omega_x$ and $\omega_z$, and the frequencies of the $\Omega$ and $\Omega_p$ drives, $\omega_r$ and $\omega_p$.
We have also set the zero energy level for the atoms to be at $\ket{u}$, so that the frequencies $\omega_d$, $\omega_s$, and $\omega_c$ are atomic transition frequencies between the respective states and the $\ket{u}$ reference state. 
The labels $u$, $d$, $s$, and $c$, are chosen to correspond to the four lowest-energy quarks where much of the group theory we utilize was originally worked out~\cite{Georgi}. 
Assuming $\Omega$ is on resonance with the $\ket{d} \leftrightarrow \ket{s}$ transition, $\omega_r = \omega_s - \omega_d$, we can move into an interaction picture which induces the rotation $\hat{\rho}_{\mathrm{sys}} = \hat{U} \hat{\rho}_1 \hat{U}^{\dagger}$ on the system's density matrix, where $\hat{U} = \exp[i \hat{H}' t / \hbar]$ with 
\begin{equation} \label{Hpr}
    \hat{H}' = \hbar \omega_d \hat{a}_x^{\dagger} \hat{a}_x + \hbar \left( \omega_p - \omega_s \right) \hat{a}_z^{\dagger} \hat{a}_z + \sum_{j = 1}^N \left[ \hbar \left( \omega_p - \omega_s \right) \hat{\sigma}_{cc}^{(j)} - \hbar \omega_d \hat{\sigma}_{dd}^{(j)} - \hbar \omega_s \hat{\sigma}_{ss}^{(j)} \right].
\end{equation}
The Hamiltonian then becomes 
\begin{equation}
    \begin{aligned}
\hat{H}_{\mathrm{sys}} = & - \hbar \Delta_x' \hat{a}_x^{\dagger} \hat{a}_x - \hbar \Delta_z' \hat{a}_z^{\dagger} \hat{a}_z - \sum_{j = 1}^N \hbar \Delta_c \hat{\sigma}_{cc}^{(j)} \\
&+ \sum_{j = 1}^N \left[ \hbar g_x \left( \hat{\sigma}_{ud}^{(j)} \hat{a}_x + \mathrm{H.c.} \right) + \hbar g_z \left( \hat{\sigma}_{cu}^{(j)} \hat{a}_z + \mathrm{H.c.} \right) + \frac{\hbar \Omega}{2} \left( \hat{\sigma}_{ds}^{(j)} + \mathrm{H.c.} \right) + \frac{\hbar \Omega_p}{2} \left( \hat{\sigma}_{sc}^{(j)} + \mathrm{H.c.} \right) \right],
    \end{aligned}
\end{equation}
with the detunings $\Delta_x' = \omega_d - \omega_x$, $\Delta_z' = \omega_p - \omega_s - \omega_z$, and $\Delta_c = \omega_p - \omega_s - \omega_c$.

Besides the coherent processes in $\hat{H}_{\mathrm{sys}}$, there are also dissipative effects stemming from the system's coupling to the environment, specifically free-space electromagnetic modes, which we model using the Lindblad superoperator
\begin{equation}
    \hat{\mathcal{D}} [\hat{O}] \hat{\rho} = \hat{O} \hat{\rho} \hat{O}^{\dagger} - \frac{1}{2} [\hat{O}^{\dagger} \hat{O} \hat{\rho} + \hat{\rho} \hat{O}^{\dagger} \hat{O}]. 
\end{equation}
As shown in Fig.~1(b) of the main text, we consider spontaneous emission from $\ket{c} \rightarrow \ket{u}$ and $\ket{c} \rightarrow \ket{s}$ at rates $\gamma_c$ and $\gamma_b$ respectively, as well as $\ket{u} \rightarrow \ket{d}$ and $\ket{u} \rightarrow \ket{s}$ at rates $\gamma_d$ and $\gamma_s$.
The cavity modes also decay into free-space at rates $\kappa_x$ and $\kappa_z$. This gives a master equation for the dynamics of the system's density matrix $\hat{\rho}_{\mathrm{sys}}$ under the Born-Markov approximation~\cite{Steck},
\begin{equation}
    \begin{aligned}
\frac{\partial \hat{\rho}_{\mathrm{sys}}}{\partial t} = & - \frac{i}{\hbar} \left[ \hat{H}_{\mathrm{sys}}, \hat{\rho}_{\mathrm{sys}} \right] + \hat{\mathcal{D}} \left[ \sqrt{\kappa_x} \hat{a}_x \right] \hat{\rho}_{\mathrm{sys}} + \hat{\mathcal{D}} \left[ \sqrt{\kappa_z} \hat{a}_z \right] \hat{\rho}_{\mathrm{sys}} \\
&+  \sum_{j = 1}^N \left( \hat{\mathcal{D}} \left[ \sqrt{\gamma_d} \hat{\sigma}_{du}^{(j)} \right] \hat{\rho}_{\mathrm{sys}} + \hat{\mathcal{D}} \left[ \sqrt{\gamma_s} \hat{\sigma}_{su}^{(j)} \right] \hat{\rho}_{\mathrm{sys}} + \hat{\mathcal{D}} \left[ \sqrt{\gamma_c} \hat{\sigma}_{uc}^{(j)} \right] \hat{\rho}_{\mathrm{sys}} + \hat{\mathcal{D}} \left[ \sqrt{\gamma_b} \hat{\sigma}_{sc}^{(j)} \right] \hat{\rho}_{\mathrm{sys}} \right).
    \end{aligned}
\end{equation}

\subsection{Adiabatic elimination of auxiliary excited state}
We now assume $\abs{\Delta_c} \gg \sqrt{N} g_z, \Omega_p$, such that the auxiliary state $\ket{c}$ can be adiabatically eliminated over the coarse-grained timescale that characterizes the relevant system dynamics. 
Assuming $\gamma_c, \gamma_b \gg \gamma_d, \gamma_s$, since $\ket{u}$ is a metastable state, we obtain an effective master equation for the reduced density matrix $\hat{\rho}_{\mathrm{eff}} = \Tr_c[\hat{\rho}_{\mathrm{sys}}]$~\cite{Reiter},
\begin{equation} \label{MasterEq_eff_cross}
    \begin{aligned}
\frac{\partial \hat{\rho}_{\mathrm{eff}}}{\partial t} = & - \frac{i}{\hbar} \left[ \hat{H}_{\mathrm{eff}}, \hat{\rho}_{\mathrm{eff}} \right] + \hat{\mathcal{D}} \left[ \sqrt{\kappa_x} \hat{a}_x \right] \hat{\rho}_{\mathrm{eff}} + \hat{\mathcal{D}} \left[ \sqrt{\kappa_z} \hat{a}_z \right] \hat{\rho}_{\mathrm{eff}} + \sum_{j = 1}^N \Big( \hat{\mathcal{D}} \left[ \sqrt{\gamma_d} \hat{\sigma}_{du}^{(j)} \right] \hat{\rho}_{\mathrm{eff}} + \hat{\mathcal{D}} \left[ \sqrt{\gamma_s} \hat{\sigma}_{su}^{(j)} \right] \hat{\rho}_{\mathrm{eff}} \\
& + \hat{\mathcal{D}} \left[ \sqrt{w} \hat{\sigma}_{us}^{(j)} + \sqrt{\tilde{w}} \hat{\sigma}_{uu}^{(j)} \hat{a}_z \right] \hat{\rho}_{\mathrm{eff}} + \hat{\mathcal{D}} \left[ \sqrt{\gamma_p} \hat{\sigma}_{ss}^{(j)}  + \sqrt{\tilde{\gamma}_p} \hat{\sigma}_{su}^{(j)} \hat{a}_z \right] \hat{\rho}_{\mathrm{eff}} \Big),
    \end{aligned}
\end{equation}
with the effective rates 
\begin{equation}
    w = \frac{\gamma_c \Omega_p^2}{4 \left[ \Delta_c - \frac{i}{2} \left( \gamma_c + \gamma_b \right) \right]^2}, \quad \gamma_p = \frac{\gamma_b \Omega_p^2}{4 \left[ \Delta_c - \frac{i}{2} \left( \gamma_c + \gamma_b \right) \right]^2},
\end{equation}
stemming from absorption from $\Omega_p$ followed by a spontaneously emitted photon from $\ket{c}$, and 
\begin{equation}
    \tilde{w} = \frac{\gamma_c g_z^2}{4 \left[ \Delta_c - \frac{i}{2} \left( \gamma_c + \gamma_b \right) \right]^2}, \quad \tilde{\gamma}_p = \frac{\gamma_b g_z^2}{4 \left[ \Delta_c - \frac{i}{2} \left( \gamma_c + \gamma_b \right) \right]^2},
\end{equation}
stemming from absorption from the $z$-cavity followed by a spontaneously emitted photon from $\ket{c}$. 
Since jump operators always come with their Hermitian conjugate in the master equation, we can ignore the phase of the complex denominator such that
\begin{equation}
    w = \frac{\gamma_c \Omega_p^2}{4 \Delta_c^2 + \left( \gamma_c + \gamma_b \right)^2}, \quad \gamma_p = \frac{\gamma_b \Omega_p^2}{4 \Delta_c^2 + \left( \gamma_c + \gamma_b \right)^2},
\end{equation}
and
\begin{equation}
    \tilde{w} = \frac{\gamma_c g_z^2}{4 \Delta_c^2 + \left( \gamma_c + \gamma_b \right)^2}, \quad \tilde{\gamma}_p = \frac{\gamma_b g_z^2}{4 \Delta_c^2 + \left( \gamma_c + \gamma_b \right)^2}.
\end{equation}
The effective Hamiltonian in Eq.~\eqref{MasterEq_eff_cross} is given by 
\begin{equation} \label{H_eff_0}
    \begin{aligned}
\hat{H}_{\mathrm{eff}} = & - \hbar \Delta_x' \hat{a}_x^{\dagger} \hat{a}_x - \hbar \Delta_z' \hat{a}_z^{\dagger} \hat{a}_z  + \sum_{j = 1}^N \Bigg[ \hbar g_x \left( \hat{\sigma}_{ud}^{(j)} \hat{a}_x + \mathrm{H.c.} \right) + \frac{\hbar \Delta_c g_z \Omega_p}{2 \Delta_c^2 + \frac{1}{2} \left( \gamma_c + \gamma_b \right)^2} \left( \hat{\sigma}_{su}^{(j)} \hat{a}_z + \mathrm{H.c.} \right) \\
&+ \frac{\hbar \Omega}{2} \left( \hat{\sigma}_{ds}^{(j)} + \mathrm{H.c.} \right) - \frac{\hbar \Delta_c \Omega_p^2}{4 \Delta_c^2 + \left( \gamma_c + \gamma_b \right)^2} \hat{\sigma}_{ss}^{(j)} - \frac{\hbar \Delta_c g_z^2}{\Delta_c^2 + \frac{1}{4} \left( \gamma_c + \gamma_b \right)^2} \hat{\sigma}_{uu}^{(j)} \hat{a}_z^{\dagger} \hat{a}_z \Bigg]. 
    \end{aligned}
\end{equation}
Here, the last two terms represent AC Stark shifts that  shift the transition frequencies of the system.
In order to keep the physics simple, we  work in the regime where all frequency shifts are small compared to the decay rates of the cavities. 
In this limit, we can incorporate the light shift on $\ket{s}$ into the frequency of the $z$-cavity $\Delta_z' \rightarrow \tilde{\Delta}_z$ and the frequency of the RF drive $\omega_r \rightarrow \tilde{\omega}_r$, while the non-linear shift changes the detunings $\Delta_x' \rightarrow \Delta_x$ and $\tilde{\Delta}_z \rightarrow \Delta_z$. 
% The exact nature of these shifts and their effect in an active clock setting is the subject of future work. 
For this paper, we will typically assume $\Delta_x \approx \Delta_z \approx 0$, which we do at the end of this section.
However, we let $\Delta_x = \tilde{\omega}_d - \omega_x$ be finite but small, $\Delta_x / \kappa_x \ll 1$, when we study the cavity pulling, where $\tilde{\omega}_d$ is the transition frequency of the gain medium (i.e., $\omega_d$ plus the Stark shift). 
With all this, we obtain the simplified Hamiltonian,
\begin{equation}
    \hat{H}_{\mathrm{eff}} = - \hbar \Delta_x \hat{a}_x^{\dagger} \hat{a}_x - \hbar \Delta_z \hat{a}_z^{\dagger} \hat{a}_z  + \sum_{j = 1}^N \left[ \hbar g_x \left( \hat{\sigma}_{ud}^{(j)} \hat{a}_x + \mathrm{H.c.} \right) + \frac{\hbar \Delta_c g_z \Omega_p}{2 \Delta_c^2 + \frac{1}{2} \left( \gamma_c + \gamma_b \right)^2} \left( \hat{\sigma}_{su}^{(j)} \hat{a}_z + \mathrm{H.c.} \right) + \frac{\hbar \Omega}{2} \left( \hat{\sigma}_{ds}^{(j)} + \mathrm{H.c.} \right) \right].
\end{equation}
Defining the collective operators $\hat{\mathcal{C}}_+ = \sum_j \hat{\sigma}_{ud}^{(j)}$, $\hat{\mathcal{P}}_+ = \sum_j \hat{\sigma}_{us}^{(j)}$, and $\hat{\mathcal{R}}_+ = \sum_j \hat{\sigma}_{ds}^{(j)}$, we can rewrite this as 
\begin{equation} \label{H_eff}
    \hat{H}_{\mathrm{eff}} = - \hbar \Delta_x \hat{a}_x^{\dagger} \hat{a}_x - \hbar \Delta_z \hat{a}_z^{\dagger} \hat{a}_z  + \hbar g_x \left( \hat{\mathcal{C}}_+ \hat{a}_x + \mathrm{H.c.} \right) + \frac{\hbar \Delta_c g_z \Omega_p}{2 \Delta_c^2 + \frac{1}{2} \left( \gamma_c + \gamma_b \right)^2} \left( \hat{\mathcal{P}}_- \hat{a}_z + \mathrm{H.c.} \right) + \frac{\hbar \Omega}{2} \left( \hat{\mathcal{R}}_+ + \mathrm{H.c.} \right).
\end{equation}

We simplify the model further in the next subsection by assuming the cavities are both in the bad-cavity regime. 
However, we can first use this to ignore some processes from the master equation in Eq.~\eqref{MasterEq_eff_cross}.
Specifically, we compare the rates of the two terms in the jump operators in the last line of Eq.~\eqref{MasterEq_eff_cross} where we find that the terms stemming from $\Omega_p$ ($w$ and $\gamma_p$) dominate those stemming from the $z$-cavity ($\tilde{w}$ and $\tilde{\gamma}_p$). 
To see this, we assume that the system is undergoing superradiant emission through the $z$-cavity such that we can approximate the photon number of the $z$-cavity as $(N g_{\mathrm{eff}} / \kappa_z)^2$, and so we associate $\hat{a}_z \sim N g_{\mathrm{eff}} / \kappa_z$. 
Using the effective coupling rate of the $z$-cavity from Eq.~\eqref{H_eff} $g_{\mathrm{eff}} = g_z \Omega_p / (2 \Delta_c)$ (assuming the unnecessary condition $\abs{\Delta_c} \gg \gamma_b, \gamma_c$ to simplify the calculation), the conditions to ignore the $\tilde{w}$ and $\tilde{\gamma}_p$ terms are given by
\begin{equation}
    \sqrt{w} \gg \frac{N \sqrt{\tilde{w}} g_z \Omega_p}{2 \abs{\Delta_c} \kappa_z}, \quad \sqrt{\gamma_p} \gg \frac{N \sqrt{\tilde{\gamma}_p} g_z \Omega_p}{2 \abs{\Delta_c} \kappa_z},
\end{equation}
which both reduce to
\begin{equation}
    \frac{\sqrt{\gamma_k} \Omega_p}{2 \abs{\Delta_c}} \gg \frac{N \sqrt{\gamma_k} g_z^2 \Omega_p}{4 \abs{\Delta_c}^2 \kappa_z}, 
\end{equation}
where $k \in \{ b, c \}$. 
Using the cooperativity of the $z$-cavity [see Eq.~\eqref{Cooperativites}], the condition becomes
\begin{equation}
    \abs{\Delta_c} \gg \frac{N \mathscr{C}_z \gamma_c}{2}.
\end{equation}
This effectively states that the detuning of the two-photon resonance has to be well outside the collective line broadening of $\ket{c}$ from the $z$-cavity.  
Having shown that it is reasonable to operate in a regime where one can ignore the $\tilde{w}$ and $\tilde{\gamma}_p$ terms, we derive the master equation
\begin{equation} %\label{MasterEq_eff}
    \begin{aligned}
\frac{\partial \hat{\rho}_{\mathrm{eff}}}{\partial t} = & - \frac{i}{\hbar} \left[ \hat{H}_{\mathrm{eff}}, \hat{\rho}_{\mathrm{eff}} \right] + \hat{\mathcal{D}} \left[ \sqrt{\kappa_x} \hat{a}_x \right] \hat{\rho}_{\mathrm{eff}} + \hat{\mathcal{D}} \left[ \sqrt{\kappa_z} \hat{a}_z \right] \hat{\rho}_{\mathrm{eff}} \\
&+  \sum_{j = 1}^N \hat{\mathcal{D}} \left[ \sqrt{\gamma_d} \hat{\sigma}_{du}^{(j)} \right] \hat{\rho}_{\mathrm{eff}} + \hat{\mathcal{D}} \left[ \sqrt{\gamma_s} \hat{\sigma}_{su}^{(j)} \right] \hat{\rho}_{\mathrm{eff}} + \hat{\mathcal{D}} \left[ \sqrt{w} \hat{\sigma}_{us}^{(j)} \right] \hat{\rho}_{\mathrm{eff}} + \hat{\mathcal{D}} \left[ \sqrt{\gamma_p} \hat{\sigma}_{ss}^{(j)} \right] \hat{\rho}_{\mathrm{eff}}.
    \end{aligned}
\end{equation}

\subsection{Adiabatic elimination of cavity fields}
Superradiant lasing is achieved in the bad-cavity regime, $\kappa_x \gg \sqrt{N} g_x, \gamma_d, \gamma_s, w, \gamma_p$ and $\kappa_z \gg \sqrt{N} \abs{\Delta_c} g_z \Omega_p / [2 \Delta_c^2 + \left( \gamma_c + \gamma_b \right)^2 / 2], \gamma_d, \gamma_s, w, \gamma_p$. 
Working in this regime allows us to adiabatically eliminate the cavity fields over a coarse-grained timescale~\cite{Jager}.
We follow the method of Ref.~\cite{Xu} to obtain the effective master equation for the reduced density matrix $\hat{\rho} = \Tr_{x,z} [\hat{\rho}_{\mathrm{eff}}]$,
\begin{equation} \label{MasterEq_nzDeltas}
    \pd{\hat{\rho}} = - \frac{i}{\hbar} \left[ \hat{H}, \hat{\rho} \right] + \hat{\mathcal{D}} \left[ \sqrt{\Gamma_c} \, \hat{\mathcal{C}}_- \right] \hat{\rho} + \hat{\mathcal{D}} \left[ \sqrt{W} \hat{\mathcal{P}}_+ \right] \hat{\rho} + \hat{\mathcal{L}}_{\mathrm{sp}} \hat{\rho}.
\end{equation}
Here, we have defined the Hamiltonian 
\begin{equation}
    \hat{H} = \frac{\hbar \Omega}{2} \left( \hat{\mathcal{R}}_+ + \hat{\mathcal{R}}_- \right) - \hbar \chi_x \hat{\mathcal{C}}_+ \hat{\mathcal{C}}_- - \hbar \chi_z \hat{\mathcal{P}}_- \hat{\mathcal{P}}_+,
\end{equation}
the single-particle Liouvillian
\begin{equation}
    \hat{\mathcal{L}}_{\mathrm{sp}} = \sum_{j = 1}^N \hat{\mathcal{D}} \left[ \sqrt{\gamma_d} \hat{\sigma}_{du}^{(j)} \right] + \hat{\mathcal{D}} \left[ \sqrt{\gamma_s} \hat{\sigma}_{su}^{(j)} \right] + \hat{\mathcal{D}} \left[ \sqrt{w} \hat{\sigma}_{us}^{(j)} \right] + \hat{\mathcal{D}} \left[ \sqrt{\gamma_p} \hat{\sigma}_{ss}^{(j)} \right],
\end{equation}
and the effective rates from the $x$-cavity,
\begin{equation}
    \Gamma_c = \frac{\kappa_x g_x^2}{4 \Delta_x^2 + \kappa_x^2}, \quad \chi_x = \frac{\Delta_x g_x^2}{4 \Delta_x^2 + \kappa_x^2},
\end{equation}
and from the $z$-cavity,
\begin{equation}
    W = \frac{\kappa_z \Delta_c^2 g_z^2 \Omega_p^2}{\left[ 2 \Delta_c^2 + \frac{1}{2} \left( \gamma_c + \gamma_b \right)^2 \right]^2 \left[ 4 \Delta_z^2 + \kappa_z^2 \right]}, \quad \chi_z = \frac{\Delta_z \Delta_c^2 g_z^2 \Omega_p^2}{\left[ 2 \Delta_c^2 + \frac{1}{2} \left( \gamma_c + \gamma_b \right)^2 \right]^2 \left[ 4 \Delta_z^2 + \kappa_z^2 \right]}.
\end{equation}
Setting $\Delta_x \approx \Delta_z \approx 0$ and defining $\hat{\mathcal{R}}_x = \left( \hat{\mathcal{R}}_+ + \hat{\mathcal{R}}_- \right) / 2$, we obtain a final master equation,
\begin{equation} \label{MasterEq}
    \hat{\mathcal{L}} \hat{\rho} \equiv \pd{\hat{\rho}} = i \Omega \left[ \hat{\rho}, \hat{\mathcal{R}}_x \right] + \hat{\mathcal{D}} \left[ \sqrt{\Gamma_c} \, \hat{\mathcal{C}}_- \right] \hat{\rho} + \hat{\mathcal{D}} \left[ \sqrt{W} \hat{\mathcal{P}}_+ \right] \hat{\rho} + \hat{\mathcal{L}}_{\mathrm{sp}} \hat{\rho},
\end{equation}
which encapsulates the model presented in Eq.~(1) of the main text.
Here, $\hat{\mathcal{L}}$ is the Liouvillian superoperator whose kernel gives the system's steady-state,
\begin{equation}
    \hat{\mathcal{L}} \hat{\rho}_{\mathrm{ss}} = 0.
\end{equation}
We can also define the  cooperativity parameters for the two cavities as
\begin{equation} \label{Cooperativites}
    \mathscr{C}_x = \frac{\Gamma_c}{\gamma_d} = \frac{g_x^2}{\gamma_d \kappa_x}, \quad \mathscr{C}_z = \frac{W}{w} = \frac{\Delta_c^2 g_z^2}{\gamma_c \kappa_z \left[ \Delta_c^2 + \frac{1}{4} \left( \gamma_c + \gamma_b \right)^2 \right]} \approx \frac{g_z^2}{\gamma_c \kappa_z}.
\end{equation}

\section{U(1) Symmetry} \label{Sec:U1}
In this section, we discuss how to reduce the Liouville space scaling for exact simulations by utilizing an additional $\mathrm{U}(1)$ symmetry of the master equation, Eq.~(1) of the main text. 
To see this additional symmetry, note that one can put an arbitrary phase on the excited state, $\ket{u} \rightarrow e^{i \theta} \ket{u}$, without changing the system's dynamics. This is due to $\ket{u}$ only entering the master equation through dissipative terms, and thus any $\ket{u}$ is always accompanied by a $\bra{u}$ in the Hermitian conjugate. 
We now explore how to reduce the system's scaling using this additional symmetry in exact master equation simulations. 

\subsection{Superoperator actions}
We focus on the fully collective case, where we manually set $\hat{\mathcal{L}}_{\mathrm{sp}} = 0$.
Using Ref.~\cite{Reilly3}, we know that the symmetric subspace of $\mathrm{SU}(3)$ can be represented by two quantum numbers $\{ r, r_3 \}$ that are the eigenvalues for the $\mathcal{R}$ transition:
\begin{equation}
    \hat{\mathcal{R}}^2 \ket{r,r_3} = r(r + 1) \ket{r,r_3}, \quad \hat{\mathcal{R}}_z \ket{r,r_3} = r_3 \ket{r,r_3}. 
\end{equation}
This means that the Hilbert space scales as $(N + 2)(N + 1) / 2 \sim N^2 / 2$, and so the Liouville space, given by $\mathrm{span}[\{ \ket{r,r_3} \otimes \ket{r',r_3'}^* \}, \, \forall r,r_3,r',r_3']$, scales as $(N + 2)^2 (N + 1)^2 / 4 \sim N^4 / 4$. 
Let us now see how the $\mathrm{U}(1)$ symmetry can reduce the scaling.

First, we work out
\begin{equation}
    \begin{aligned}
\hat{\mathcal{C}}_+ \ket{r,r_3} &= \sqrt{(r + r_3)(N - 2 r + 1)} \ket{r - \frac{1}{2},r_3 - \frac{1}{2}}, \quad \hat{\mathcal{C}}_- \ket{r,r_3} = \sqrt{(N - 2 r)(r + r_3 + 1)} \ket{r + \frac{1}{2},r_3 + \frac{1}{2}}, \\
\hat{\mathcal{P}}_+ \ket{r,r_3} &= \sqrt{(r - r_3)(N - 2 r + 1)} \ket{r - \frac{1}{2},r_3 + \frac{1}{2}}, \quad \hat{\mathcal{P}}_- \ket{r,r_3} = \sqrt{(N - 2 r)(r - r_3 + 1)} \ket{r + \frac{1}{2},r_3 - \frac{1}{2}}, \\
\hat{\mathcal{R}}_+ \ket{r,r_3} &= \sqrt{(r - r_3)(r + r_3 + 1)} \ket{r,r_3 + 1}, \qquad \qquad \hat{\mathcal{R}}_- \ket{r,r_3} = \sqrt{(r + r_3)(r - r_3 + 1)} \ket{r,r_3 - 1}.
    \end{aligned}
\end{equation}
We now map the density matrix into Liouville space~\cite{Gyamfi},
\begin{equation}
    \hat{\rho} = \sum_{r,r_3,r',r_3'} c_{r,r_3,r',r_3'} \op{r,r_3}{r',r_3'} \longrightarrow \ket{\hat{\rho} \rangle} = \sum_{r,r_3,r',r_3'} c_{r,r_3,r',r_3'} \ket{r,r_3} \otimes \ket{r',r_3'}^*,
\end{equation}
with the superoperator mappings
\begin{equation}
    \begin{aligned}
\hat{O}_1 \hat{\rho} \longrightarrow & \left( \hat{O}_1 \otimes \hat{\mathbb{I}} \right) \ket{\hat{\rho} \rangle}, \\ 
\hat{\rho} \hat{O}_2 \longrightarrow & \left( \hat{\mathbb{I}} \otimes \hat{O}_2^T \right) \ket{\hat{\rho} \rangle}, \\ 
\hat{O}_1 \hat{\rho} \hat{O}_2 \longrightarrow & \left( \hat{O}_1 \otimes \hat{O}_2^T \right) \ket{\hat{\rho} \rangle},
    \end{aligned}
\end{equation}
for any operators $\hat{O}_1$ and $\hat{O}_2$. 
To perform a trace, we form the identity vector $\ket{\mathbb{I} \rangle} = \sum_{r,r_3} \ket{r,r_3} \otimes \ket{r,r_3}^*$ and take 
\begin{equation}
    \expval{\hat{O}} = \Tr[ \hat{O} ] = \bra{\langle \mathbb{I}} \hat{O} \otimes \hat{\mathbb{I}} \ket{\hat{\rho} \rangle},
\end{equation}
such that the normalization condition is $\ip{\langle \mathbb{I}}{\hat{\rho} \rangle} = 1$. 
The fully collective master equation, acting on an arbitrary matrix element, consists of eight superoperators:
\begin{subequations} \label{RelevantSuperOps}
\begin{equation}
    \hat{\mathcal{R}}_x \op{r,r_3}{r',r_3'} = \frac{1}{2} \left[ \sqrt{(r - r_3)(r + r_3 + 1)} \op{r,r_3 + 1}{r',r_3'} + \sqrt{(r + r_3)(r - r_3 + 1)} \op{r,r_3 - 1}{r',r_3'} \right],
\end{equation}
\begin{equation}
    \op{r,r_3}{r',r_3'} \hat{\mathcal{R}}_x = \frac{1}{2} \left[ \left( \sqrt{(r' + r_3')(r' - r_3' + 1)} \right)^* \op{r,r_3}{r',r_3' - 1} + \left( \sqrt{(r' - r_3')(r' + r_3' + 1)} \right)^* \op{r,r_3}{r',r_3' + 1} \right],
\end{equation}
\begin{equation}
    \hat{\mathcal{C}}_- \op{r,r_3}{r',r_3'} \hat{\mathcal{C}}_+ = \sqrt{(N - 2 r)(r + r_3 + 1)} \left( \sqrt{(N - 2 r')(r' + r'_3 + 1)} \right)^* \op{r + \frac{1}{2},r_3 + \frac{1}{2}}{r' + \frac{1}{2},r_3' + \frac{1}{2}},
\end{equation}
\begin{equation}
    \hat{\mathcal{C}}_+ \hat{\mathcal{C}}_- \op{r,r_3}{r',r_3'} = (N - 2 r)(r + r_3 + 1) \op{r,r_3}{r',r_3'},
\end{equation}
\begin{equation}
    \op{r,r_3}{r',r_3'} \hat{\mathcal{C}}_+ \hat{\mathcal{C}}_- = \left[ (N - 2 r')(r' + r_3' + 1) \right]^* \op{r,r_3}{r',r_3'},
\end{equation}
\begin{equation}
    \hat{\mathcal{P}}_+ \op{r,r_3}{r',r_3'} \hat{\mathcal{P}}_- = \sqrt{(r - r_3)(N - 2 r + 1)} \left( \sqrt{(r' - r_3')(N - 2 r' + 1)} \right)^* \op{r - \frac{1}{2},r_3 + \frac{1}{2}}{r' - \frac{1}{2},r_3' + \frac{1}{2}},
\end{equation}
\begin{equation}
    \hat{\mathcal{P}}_- \hat{\mathcal{P}}_+ \op{r,r_3}{r',r_3'} = (r - r_3)(N - 2 r + 1) \op{r,r_3}{r',r_3'},
\end{equation}
\begin{equation}
    \op{r,r_3}{r',r_3'} \hat{\mathcal{P}}_- \hat{\mathcal{P}}_+ = \left[ (r' - r_3')(N - 2 r' + 1) \right]^* \op{r,r_3}{r',r_3'}.
\end{equation}
\end{subequations}
Thus, we see that the $\mathrm{U}(1)$ symmetry emerges from the property that the Liouvillian always changes $r$ and $r'$ symmetrically. 
Since we typically begin in the state 
\begin{equation}
    \hat{\rho}_0 = \op{d}{d}^{\otimes N} = \op{r = N/2, r_3 = N / 2}{r' = N/2, r_3' = N / 2},
\end{equation}
this means that only elements of our density matrix with $r' = r$ will ever become non-zero during the evolution and so we can reduce our state space. 
Therefore, we obtain bounds on our quantum numbers,
\begin{equation}
    \begin{aligned}
& r \in \{ 0, 1/2, \ldots, N/2 \}, \\
& r_3 \in \{ -r, -r + 1, \ldots, r - 1, r \}, \\
& r_3' \in \{ -r, -r + 1, \ldots, r - 1, r \},
    \end{aligned}
\end{equation}
which give a state space scaling of 
\begin{equation} \label{U1_DiagScaling}
    d_N = \sum_{j = 0}^N (j + 1)^2 = \frac{N^3}{3} + \frac{3 N^2}{2} + \frac{13 N}{6} + 1 \sim \frac{N^3}{3}.
\end{equation}
We also find that all the expressions in Eq.~\eqref{RelevantSuperOps} are real, and so they can be written in Liouville space, with the notation $\op{r,r_3}{r,r_3'} \rightarrow \ket{r,r_3,r_3' \rangle}$, as
\begin{subequations}
\begin{equation}
    \hat{\mathcal{R}}_x \op{r,r_3}{r,r_3'} \rightarrow \frac{1}{2} \left[ \sqrt{(r - r_3)(r + r_3 + 1)} \ket{r,r_3 + 1,r_3' \rangle} + \sqrt{(r + r_3)(r - r_3 + 1)} \ket{r,r_3 - 1,r_3' \rangle} \right],
\end{equation}
\begin{equation}
    \op{r,r_3}{r,r_3'} \hat{\mathcal{R}}_x \rightarrow \frac{1}{2} \left[ \sqrt{(r + r_3')(r - r_3' + 1)} \ket{r,r_3,r_3' - 1 \rangle} + \sqrt{(r - r_3')(r + r_3' + 1)} \ket{r,r_3,r_3' + 1 \rangle} \right],
\end{equation}
\begin{equation}
    \hat{\mathcal{C}}_- \op{r,r_3}{r,r_3'} \hat{\mathcal{C}}_+ \rightarrow (N - 2 r) \sqrt{(r + r_3 + 1) (r + r'_3 + 1)} \ket{r + \frac{1}{2},r_3 + \frac{1}{2},r_3' + \frac{1}{2} \Bigg\rangle},
\end{equation}
\begin{equation}
    \hat{\mathcal{C}}_+ \hat{\mathcal{C}}_- \op{r,r_3}{r,r_3'} \rightarrow (N - 2 r)(r + r_3 + 1) \ket{r,r_3,r_3' \rangle},
\end{equation}
\begin{equation}
    \op{r,r_3}{r,r_3'} \hat{\mathcal{C}}_+ \hat{\mathcal{C}}_- \rightarrow (N - 2 r)(r + r_3' + 1) \ket{r,r_3,r_3' \rangle},
\end{equation}
\begin{equation}
    \hat{\mathcal{P}}_+ \op{r,r_3}{r,r_3'} \hat{\mathcal{P}}_- \rightarrow (N - 2 r + 1) \sqrt{(r - r_3)(r - r_3')} \ket{r - \frac{1}{2},r_3 + \frac{1}{2},r_3' + \frac{1}{2} \Bigg\rangle},
\end{equation}
\begin{equation}
    \hat{\mathcal{P}}_- \hat{\mathcal{P}}_+ \op{r,r_3}{r,r_3'} \rightarrow (r - r_3)(N - 2 r + 1) \ket{r,r_3,r_3' \rangle},
\end{equation}
\begin{equation}
    \op{r,r_3}{r,r_3'} \hat{\mathcal{P}}_- \hat{\mathcal{P}}_+ \rightarrow (r - r_3')(N - 2 r + 1) \ket{r,r_3,r_3' \rangle}.
\end{equation}
\end{subequations}

\subsection{Linewidth}
While the basis in the previous subsection is sufficient to calculate observables such as state populations and the intensities of the respective cavity modes, it is not sufficient for calculating the linewidth of the output light field. 
This is because the linewidth describes the decay of coherences that include terms such as 
\begin{equation}
    \op{\alpha, \beta}{\alpha - 1, \beta + 1} = \op{r,r_3}{r - \frac{1}{2}, r_3 - \frac{1}{2}},
\end{equation}
where $\alpha$ and $\beta$ are the number of atoms in $\ket{u}$ and $\ket{d}$, respectively. 
We can follow a more general method where we can write the dynamics in $k$-sectors defined by~\cite{Reilly3} $\ket{k,r,r_3,r_3' \rangle} \equiv \ket{r,r_3} \otimes \ket{r - k/2, r_3' \rangle}$,
with the bounds
\begin{equation}
    \begin{aligned}
& k \in \{ 0, 1 \}, \\
& r \in \{ 0, 1/2, \ldots, N/2 \}, \\
& r_3 \in \{ -r, -r + 1, \ldots, r - 1, r \}, \\
& r_3' \in \{ -r + k/2, -r + k/2 + 1, \ldots, r - k/2 - 1, r - k/2 \}.
    \end{aligned}
\end{equation}
The dynamics conserve $k$ and only modify the quantum numbers $r$, $r_3$, and $r_3'$. 
We can describe the action of all superoperators in each $k$-sector where we find
\begin{subequations}
\begin{equation}
    \hathat{\mathbb{L}} [\hat{\mathcal{R}}_+] \ket{k,r,r_3,r_3' \rangle} = \sqrt{(r - r_3)(r + r_3 + 1)} \ket{k,r,r_3 + 1,r_3' \rangle},
\end{equation}
\begin{equation}
    \hathat{\mathbb{R}} [\hat{\mathcal{R}}_+] \ket{k,r,r_3,r_3' \rangle} = \sqrt{\left( r - \frac{k}{2} + r_3' \right) \left( r - \frac{k}{2} - r_3' + 1 \right)} \ket{k,r,r_3,r_3' - 1 \rangle},
\end{equation}
\begin{equation}
    \hathat{\mathbb{L}} [\hat{\mathcal{R}}_-] \ket{k,r,r_3,r_3' \rangle} = \sqrt{(r + r_3)(r - r_3 + 1)} \ket{k,r,r_3 - 1,r_3' \rangle},
\end{equation}
\begin{equation}
    \hathat{\mathbb{R}} [\hat{\mathcal{R}}_-] \ket{k,r,r_3,r_3' \rangle} = \sqrt{\left( r - \frac{k}{2} - r_3' \right) \left( r - \frac{k}{2} + r_3' + 1 \right)} \ket{k,r,r_3,r_3' + 1 \rangle},
\end{equation}
\begin{equation}
    \hathat{\mathbb{B}}[\hat{\mathcal{C}}_-, \hat{\mathcal{C}}_+] \ket{k,r,r_3,r_3' \rangle} = \sqrt{(N - 2 r) (r + r_3 + 1) (N - 2 r + k) \left( r - \frac{k}{2} + r'_3 + 1 \right)} \ket{k,r + \frac{1}{2},r_3 + \frac{1}{2},r_3' + \frac{1}{2} \Bigg\rangle},
\end{equation}
\begin{equation}
    \hathat{\mathbb{L}} [\hat{\mathcal{C}}_+ \hat{\mathcal{C}}_-] \ket{k,r,r_3,r_3' \rangle} = (N - 2 r)(r + r_3 + 1) \ket{k,r,r_3,r_3' \rangle},
\end{equation}
\begin{equation}
    \hathat{\mathbb{R}} [\hat{\mathcal{C}}_+ \hat{\mathcal{C}}_-] \ket{k,r,r_3,r_3' \rangle} = (N - 2 r + k) \left( r - \frac{k}{2} + r_3' + 1 \right) \ket{k,r,r_3,r_3' \rangle},
\end{equation}
\begin{equation}
    \hathat{\mathbb{B}} [\hat{\mathcal{P}}_+, \hat{\mathcal{P}}_-] \ket{k,r,r_3,r_3' \rangle} = \sqrt{(r - r_3) (N - 2 r + 1) \left( r - \frac{k}{2} - r_3' \right) (N - 2 r + k + 1)} \ket{k,r - \frac{1}{2},r_3 + \frac{1}{2},r_3' + \frac{1}{2} \Bigg\rangle},
\end{equation}
\begin{equation}
    \hathat{\mathbb{L}} [\hat{\mathcal{P}}_- \hat{\mathcal{P}}_+] \ket{k,r,r_3,r_3' \rangle} = (r - r_3)(N - 2 r + 1) \ket{k,r,r_3,r_3' \rangle},
\end{equation}
\begin{equation}
    \hathat{\mathbb{L}} [\hat{\mathcal{P}}_- \hat{\mathcal{P}}_+] \ket{k,r,r_3,r_3' \rangle} = \left( r - \frac{k}{2} - r_3' \right)(N - 2 r + k + 1) \ket{k,r,r_3,r_3' \rangle},
\end{equation}
\end{subequations}
with 
\begin{equation}
    \hathat{\mathbb{L}} [\hat{O}] = \hat{O} \otimes \hat{\mathbb{I}}, \quad  \hathat{\mathbb{R}} [\hat{O}] = \hat{\mathbb{I}} \otimes \hat{O}^T, \quad \hathat{\mathbb{B}} [\hat{O}_1, \hat{O}_2] =  \hat{O}_1 \otimes \hat{O}_2^T.
\end{equation}
The specific examples of this $k$-sector decomposition that concern us here are $k = 0$ which is the sector that includes the stationary state and $k = 1$ which describes the decay of coherences and therefore the linewidth of the laser. 
The addition of the $k = 1$ sector adds an additional
\begin{equation}
    \sum_{j = 0}^N j (j + 1) = \frac{N (N + 1) (N + 2)}{3}
\end{equation}
states to the basis which, along with Eq.~\eqref{U1_DiagScaling}, gives a total scaling of 
\begin{equation}
    d_N = \frac{2N^3}{3} + \frac{5 N^2}{2} + \frac{17 N}{6} + 1.
\end{equation}

\section{Candidate Transitions}
In this section, we look at the clock transitions of candidate atomic species to put characteristic values to our results. 
Common atomic species that have garnered attention for optical lattice atomic clocks are ${}^{87} \mathrm{Sr}$~\cite{Bothwell,Robinson} and ${}^{171} \mathrm{Yb}$~\cite{Hinkley,PedrozoPenafiel}, which both have a $\mathcal{O}(\mathrm{mHz})$ linewidth on the ${}^1 S_0 \leftrightarrow {}^3 P_0$ transition. 
The latter is particularly interesting for the SU(3) model as it has a nuclear spin of $I = 1 / 2$, and so two ground states. 
However, they both require the pumping scheme to become a three-photon transition~\cite{Meiser}, which may lead to extra decoherence effects. 
To avoid this, we instead look at species whose clock transition is between $S$ and $D$ orbitals with an auxiliary $P$ excited state to mediate pumping. 
Some interesting examples of this are barium~\cite{Dzuba2,Dammalapati,Yu,Yu2,Yu3,Yu4}, copper~\cite{Dzuba,Bothwell2}, and gold~\cite{Dzuba}, which all have $\mathcal{O}(\mathrm{mHz})$ linewidth clock transitions. 

\begin{figure}
    \centerline{\includegraphics[width=0.75\linewidth]{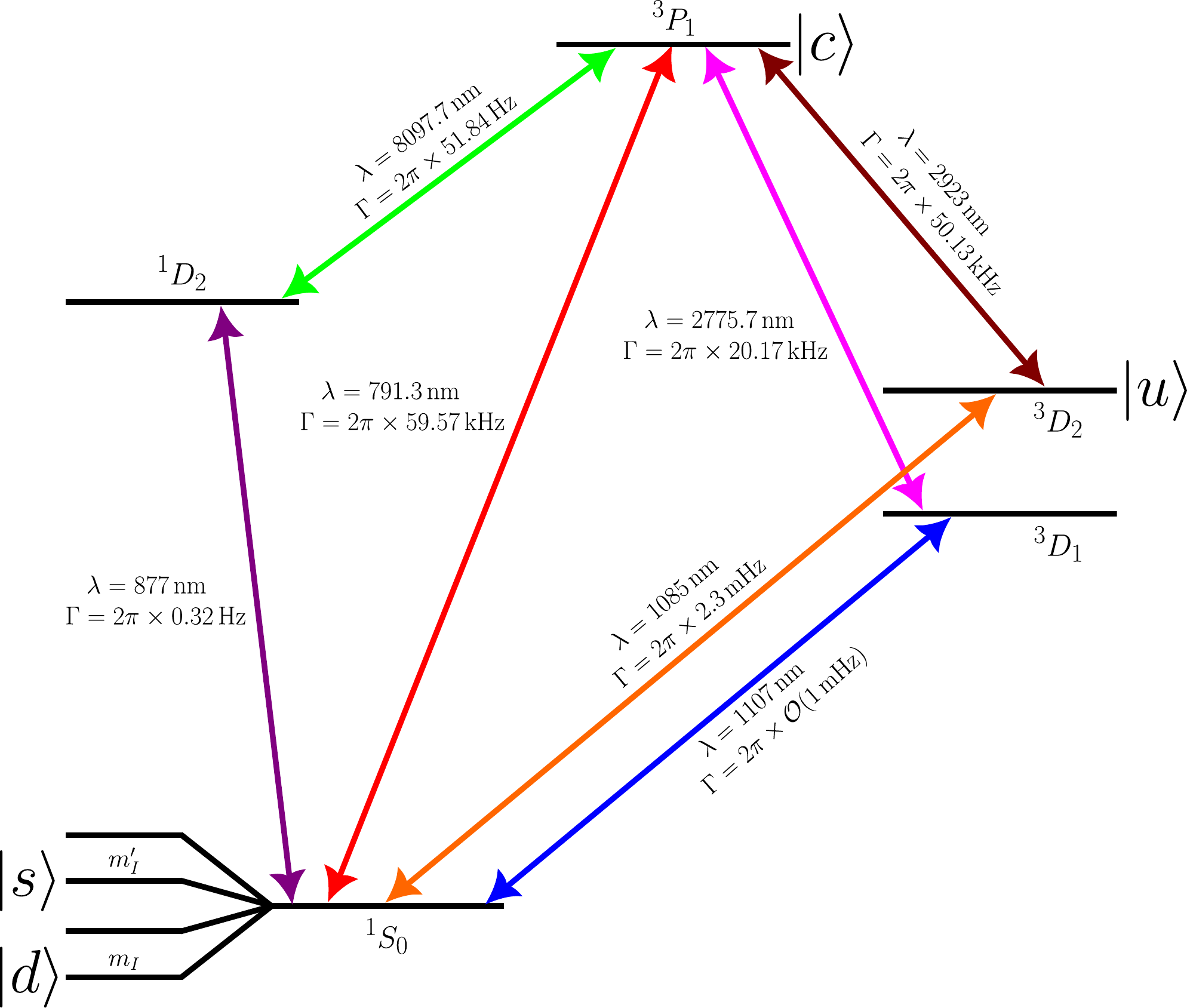}}
    \caption{Schematic diagram of the relevant states of barium.
    The wavelength $\lambda$ and decay rate $\Gamma$ of each transition, taken from Refs.~\cite{Dzuba2,Dammalapati}, is shown.}
    \label{BariumSchematic}
\end{figure}
We consider ${}^{135} \mathrm{Ba}$ which has $I = 3 / 2$, and so hyperfine structure in the ground state. 
Barium has been cooled and trapped experimentally~\cite{De,De2}. 
The relevant states for the SU(3) superradiant laser are shown in Fig.~\ref{BariumSchematic} with their transition wavelength $\lambda$ and linewidth $\Gamma$. 
There are three optical clock transitions~\cite{Yu,Yu2,Yu3}, but we focus on the $1085 \, \mathrm{nm}$ ${}^{1} S_0 \leftrightarrow {}^{3} D_2$ transition as it has been the most studied. 
We therefore choose $\ket{d} = \ket{{}^1 S_0; m_I}$, $\ket{s} = \ket{{}^1 S_0; m_I'}$, $\ket{c} = \ket{{}^3 P_1}$, and $\ket{u} = \ket{{}^3 D_2}$, where $m_I$ and $m_I'$ are different nuclear spin states, as shown in Fig.~\ref{BariumSchematic}. 
In principle, extra repump lasers would need to be added to remove population that decays into unwanted hyperfine states, but we will not consider such schemes here. 
To determine parameters, we first assume $\gamma_d \approx \gamma_s \approx 2.3 \, \mathrm{mHz}$ on the clock transition~\cite{Dzuba2}. 
Then, we assume that the cooperativity parameters of the two cavities [see Eq.~\eqref{Cooperativites}] are $\mathscr{C}_x = \mathscr{C}_z = 0.1$ which is experimentally feasible~\cite{Norcia,Norcia2}. 
Lastly, we use the branching ratio between ${}^{3} P_1 \rightarrow {}^{3} D_2$ and ${}^{3} P_1 \rightarrow {}^{1} S_0$ to approximate $\gamma_p / w \approx 1.2$, which is a particularly advantageous property of barium. 
Thus, the choice of $W = 15 \Gamma_c$ determines all of the model's parameters except the RF Rabi frequency $\Omega$, which will be $\mathcal{O} (N \Gamma_c)$ in the superradiant lasing regime. 
We summarize all of the parameters in Table~\ref{tb:BaParams}.

\begin{table}[ht]
    \centering
    \begin{tabular}{|c||c|}
\hline
Parameter & Angular Frequency \\
\hline
\hline
    $\gamma_d$ & $2.3 \, \mathrm{mHz}$ \\
\hline 
    $\gamma_s$ & $2.3 \, \mathrm{mHz}$ \\
\hline
    $w$ & $34.5 \, \mathrm{mHz}$ \\
\hline
    $\gamma_p$ & $41.5 \, \mathrm{mHz}$ \\
\hline
    $\Gamma_c$ & $0.23 \, \mathrm{mHz}$ \\
\hline
    $W$ & $3.45 \, \mathrm{mHz}$ \\
\hline
    $\Omega$ & $\mathcal{O}(N \Gamma_c)$ \\
\hline
    \end{tabular}
    \caption{Summary of parameters for barium as discussed in the text.}
    \label{tb:BaParams}
\end{table}

\section{Mean-Field Analysis} 
In this section, we perform a mean-field approximation which is used to predict the mean-field quantities of the threshold, linewidth, and pulling coefficient for very large atom numbers $N \gg 100$.

\subsection{Dynamical description}
We separate the dynamics of the atoms into parts that include collective and single-particle processes.

\paragraph{Collective Dynamics:}
We begin with the Heisenberg-Langevin equations describing the collective dynamics that can be written as
\begin{equation}
    \left[\pd{\hat{O}}\right]_{\mathrm{coll.}} = \hat{\mathcal{L}}^{\dagger} \hat{O} + \hat{\mathcal{F}} \left[ \sqrt{W} \hat{P}_+ \right] \hat{O} + \hat{\mathcal{F}} \left[ \sqrt{\Gamma_c} \hat{C}_- \right] \hat{O},
    \end{equation}
with
\begin{equation}
      \hat{\mathcal{L}}^{\dagger} \hat{O} \equiv - i \Omega \left[ \hat{O}, \hat{\mathcal{R}}_x \right] + W \left( \hat{\mathcal{P}}_- \hat{O} \hat{\mathcal{P}}_+ - \frac{1}{2} \left\{ \hat{\mathcal{P}}_- \hat{\mathcal{P}}_+, \hat{O} \right\} \right) + \Gamma_c \left( \hat{\mathcal{C}}_+ \hat{O} \hat{\mathcal{C}}_- - \frac{1}{2} \left\{ \hat{\mathcal{C}}_+ \hat{\mathcal{C}}_-, \hat{O} \right\} \right),
\end{equation}
and the Langevin noise operator for jump operator $\sqrt{R} \hat{J}$ acting on operator $\hat{O}$~\cite{Gardiner},
\begin{equation}
    \hat{\mathcal{F}} \left[ \sqrt{R} \hat{J} \right] \hat{O} = \sqrt{R} \left( \left[ \hat{O}, \hat{J}^{\dagger} \right] \hat{\xi}_R + \hat{\xi}_R^{\dagger} \left[ \hat{J}, \hat{O} \right] \right).
\end{equation}
Here, $\hat{\xi}_R$ is the vacuum quantum white noise operator for the respective decoherence process with $\expval{\hat{\xi}_R} = 0$, $\expval{\hat{\xi}_{R'}^{\dagger} (t') \hat{\xi}_R (t)} = 0$, and $\expval{\hat{\xi}_R (t) \hat{\xi}_{R'}^{\dagger} (t')} = \delta_{R,R'} \delta(t - t')$ (for $R = W, \Gamma_c$).
Using the commutation relations presented in Table~\ref{tb:CommutationRelations}, we calculate
\begin{equation} \label{MF_Coherences}
    \begin{aligned}
\left[ \pd{\hat{\mathcal{C}}_-} \right]_{\mathrm{coll.}} &= \frac{i \Omega}{2} \hat{\mathcal{P}}_- + \frac{W}{2} \hat{\mathcal{P}}_- \hat{\mathcal{R}}_+ + \Gamma_c \hat{\mathcal{C}}_z \hat{\mathcal{C}}_- - \sqrt{W} \hat{\xi}_W^{\dagger} \hat{\mathcal{R}}_+ - 2 \sqrt{\Gamma_c} \hat{\mathcal{C}}_z \hat{\xi}_{\Gamma_c}, \\
\left[ \pd{\hat{\mathcal{P}}_-} \right]_{\mathrm{coll.}} &= \frac{i \Omega}{2} \hat{\mathcal{C}}_- - W \hat{\mathcal{P}}_- \hat{\mathcal{P}}_z - \frac{\Gamma_c}{2} \hat{\mathcal{R}}_- \hat{\mathcal{C}}_- + 2 \sqrt{W} \hat{\xi}_W^{\dagger} \hat{\mathcal{P}}_z + \sqrt{\Gamma_c} \hat{\mathcal{R}}_- \hat{\xi}_{\Gamma_c}, \\
\left[ \pd{\hat{\mathcal{R}}_-} \right]_{\mathrm{coll.}} &= i \Omega \hat{\mathcal{R}}_z - \frac{W}{2} \hat{\mathcal{P}}_- \hat{\mathcal{C}}_+ + \frac{\Gamma_c}{2} \hat{\mathcal{C}}_+ \hat{\mathcal{P}}_- + \sqrt{W} \hat{\xi}_W^{\dagger} \hat{\mathcal{C}}_+ - \sqrt{\Gamma_c} \hat{\xi}_{\Gamma_c}^{\dagger} \hat{\mathcal{P}}_-.
    \end{aligned}
\end{equation}
\begin{table*}[ht]
    \centering
    \begin{tabular}{|c||c|c|c||c|c|c||c|c|c|}
    \hline
        & & & & & & & & & \\
        $[\hat{A}_{\downarrow}, \hat{B}_{\rightarrow}] = \hat{O}$ & $\hat{\mathcal{C}}_+$ & $\hat{\mathcal{C}}_-$ & $\hat{\mathcal{C}}_z$ & $\hat{\mathcal{P}}_+$ & $\hat{\mathcal{P}}_-$ & $\hat{\mathcal{P}}_z$ & $\hat{\mathcal{R}}_+$ & $\hat{\mathcal{R}}_-$ & $\hat{\mathcal{R}}_z$ \\
        & & & & & & & & & \\
    \hline
    \hline
        $\hat{\mathcal{C}}_+$ & $0$ & $2 \hat{\mathcal{C}}_z$ & $- \hat{\mathcal{C}}_+$ & $0$ & $- \hat{\mathcal{R}}_-$ & $- \frac{1}{2} \hat{\mathcal{C}}_+$ & $\hat{\mathcal{P}}_+$ & $0$ & $\frac{1}{2} \hat{\mathcal{C}}_+$ \\
    \hline 
        $\hat{\mathcal{C}}_-$ & $- 2 \hat{\mathcal{C}}_z$ & $0$ & $\hat{\mathcal{C}}_-$ & $\hat{\mathcal{R}}_+$ & $0$ & $\frac{1}{2} \hat{\mathcal{C}}_-$ & $0$ & $- \hat{\mathcal{P}}_-$ & $- \frac{1}{2} \hat{\mathcal{C}}_-$ \\
    \hline
        $\hat{\mathcal{C}}_z$ & $\hat{\mathcal{C}}_+$ & $- \hat{\mathcal{C}}_-$ & $0$ & $\frac{1}{2} \hat{\mathcal{P}}_+$ & $- \frac{1}{2} \hat{\mathcal{P}}_-$ & $0$ & $- \frac{1}{2} \hat{\mathcal{R}}_+$ & $\frac{1}{2} \hat{\mathcal{R}}_-$ & $0$ \\
    \hline
    \hline
        $\hat{\mathcal{P}}_+$ & $0$ & $- \hat{\mathcal{R}}_+$ & $- \frac{1}{2} \hat{\mathcal{P}}_+$ & $0$ & $2 \hat{\mathcal{P}}_z$ & $- \hat{\mathcal{P}}_+$ & $0$ & $\hat{\mathcal{C}}_+$ & $- \frac{1}{2} \hat{\mathcal{P}}_+$ \\
    \hline
        $\hat{\mathcal{P}}_-$ & $\hat{\mathcal{R}}_-$ & $0$ & $\frac{1}{2} \hat{\mathcal{P}}_-$ & $-2 \hat{\mathcal{P}}_z$ & $0$ & $\hat{\mathcal{P}}_-$ & $- \hat{\mathcal{C}}_-$ & $0$ & $\frac{1}{2} \hat{\mathcal{P}}_-$ \\
    \hline
        $\hat{\mathcal{P}}_z$ & $\frac{1}{2} \hat{\mathcal{C}}_+$ & $- \frac{1}{2} \hat{\mathcal{C}}_-$ & $0$ & $\hat{\mathcal{P}}_+$ & $- \hat{\mathcal{P}}_-$ & $0$ & $\frac{1}{2} \hat{\mathcal{R}}_+$ & $- \frac{1}{2} \hat{\mathcal{R}}_-$ & $0$ \\
    \hline
        \hline
        $\hat{\mathcal{R}}_+$ & $- \hat{\mathcal{P}}_+$ & $0$ & $\frac{1}{2} \hat{\mathcal{R}}_+$ & $0$ & $\hat{\mathcal{C}}_-$ & $- \frac{1}{2} \hat{\mathcal{R}}_+$ & $0$ & $2 \hat{\mathcal{R}}_z$ & $- \hat{\mathcal{R}}_+$ \\
    \hline
        $\hat{\mathcal{R}}_-$ & $0$ & $\hat{\mathcal{P}}_-$ & $- \frac{1}{2} \hat{\mathcal{R}}_-$ & $- \hat{\mathcal{C}}_+$ & $0$ & $\frac{1}{2} \hat{\mathcal{R}}_-$ & $- 2 \hat{\mathcal{R}}_z$ & $0$ & $\hat{\mathcal{R}}_-$ \\
    \hline
        $\hat{\mathcal{R}}_z$ & $- \frac{1}{2} \hat{\mathcal{C}}_+$ & $\frac{1}{2} \hat{\mathcal{C}}_-$ & $0$ & $\frac{1}{2} \hat{\mathcal{P}}_+$ & $- \frac{1}{2} \hat{\mathcal{P}}_-$ & $0$ & $\hat{\mathcal{R}}_+$ & $- \hat{\mathcal{R}}_-$ & $0$ \\
    \hline
    \end{tabular}
    \caption{Relevant commutation relations, where the first column represents the $\hat{A}$ while the first row represents the $\hat{B}$ in $[\hat{A},\hat{B}] = \hat{O}$.}
    \label{tb:CommutationRelations}
\end{table*}
Normalizing by $N$, we introduce
\begin{equation}
    \begin{aligned}
\hat{c}_- &= \hat{\mathcal{C}}_- / N, \quad \hat{p}_- = \hat{\mathcal{P}}_- / N, \quad \hat{r}_- = \hat{\mathcal{R}}_- / N, \\
\hat{c}_z &= 2 \hat{\mathcal{C}}_z / N, \quad \hat{p}_z = 2 \hat{\mathcal{P}}_z / N, \quad \hat{r}_z = 2 \hat{\mathcal{R}}_z / N,
    \end{aligned}
\end{equation}
and rewrite the Heisenberg-Langevin equations,
\begin{equation}
    \begin{aligned}
\left[ \pd{\hat{c}_-} \right]_{\mathrm{coll.}} &= \frac{i \Omega}{2} \hat{p}_- + \frac{N W}{2} \hat{p}_- \hat{r}_+ + \frac{N \Gamma_c}{2} \hat{c}_z \hat{c}_- - \sqrt{W} \hat{\xi}_W^{\dagger} \hat{r}_+ - \sqrt{\Gamma_c} \hat{c}_z \hat{\xi}_{\Gamma_c}, \\
\left[ \pd{\hat{p}_-} \right]_{\mathrm{coll.}} &= \frac{i \Omega}{2} \hat{c}_- - \frac{N W}{2} \hat{p}_- \hat{p}_z - \frac{N \Gamma_c}{2} \hat{r}_- \hat{c}_- + \sqrt{W} \hat{\xi}_W^{\dagger} \hat{p}_z + \sqrt{\Gamma_c} \hat{r}_- \hat{\xi}_{\Gamma_c}, \\
\left[ \pd{\hat{r}_-} \right]_{\mathrm{coll.}} &= \frac{i \Omega}{2} \hat{r}_z - \frac{N W}{2} \hat{p}_- \hat{c}_+ + \frac{N \Gamma_c}{2} \hat{c}_+ \hat{p}_- + \sqrt{W} \hat{\xi}_W^{\dagger} \hat{c}_+ - \sqrt{\Gamma_c} \hat{\xi}_{\Gamma_c}^{\dagger} \hat{p}_-.
    \end{aligned}
\end{equation}

\paragraph{Single-particle dynamics:}
As a next step, we collect all of the single-particle processes. 
In particular, we include effects from the single-particle Liouvillian
\begin{equation}
    \hat{\mathcal{L}}_{\mathrm{sp}} = \sum_{j = 1}^N \left(\hat{\mathcal{D}} \left[ \sqrt{\gamma_d} \hat{\sigma}_{du}^{(j)} \right] + \hat{\mathcal{D}} \left[ \sqrt{\gamma_s} \hat{\sigma}_{su}^{(j)} \right] + \hat{\mathcal{D}} \left[ \sqrt{w} \hat{\sigma}_{us}^{(j)} \right] + \hat{\mathcal{D}} \left[ \sqrt{\gamma_p} \hat{\sigma}_{ss}^{(j)} \right]\right).
\end{equation}
We define $\hat{c}_j = \hat{\sigma}_{du}^{(j)}$, $\hat{p}_j = \hat{\sigma}_{su}^{(j)}$, and $\hat{r}_j = \hat{\sigma}_{sd}^{(j)}$, and use
\begin{equation}
    \left[ \pd{\hat{O}} \right]_{\mathrm{sing.}} = \hat{\mathcal{L}}_{\mathrm{sp}}^{\dagger} \hat{O} + \sum_{j = 1}^N \left( \hat{\mathcal{F}} \left[ \sqrt{\gamma_d} \hat{c}_j \right] + \hat{\mathcal{F}} \left[ \sqrt{\gamma_s} \hat{p}_j \right] + \hat{\mathcal{F}} \left[ \sqrt{w} \hat{p}_j^{\dagger} \right] + \hat{\mathcal{F}} \left[ \sqrt{\gamma_p} \hat{p}_j \hat{p}_j^{\dagger} \right] \right) \hat{O},
\end{equation}
with
\begin{equation}
    \begin{aligned}
\hat{\mathcal{L}}_{\mathrm{sp}}^{\dagger} \hat{O} = & - \frac{\gamma_d}{2} \sum_{j = 1}^N \left( \hat{c}_j^{\dagger} \left[ \hat{c}_j, \hat{O} \right] + \left[ \hat{O}, \hat{c}_j^{\dagger} \right] \hat{c}_j \right) - \frac{\gamma_s}{2} \sum_{j = 1}^N \left( \hat{p}_j^{\dagger} \left[ \hat{p}_j, \hat{O} \right] + \left[ \hat{O}, \hat{p}_j^{\dagger} \right] \hat{p}_j \right) \\
& - \frac{w}{2} \sum_{j = 1}^N \left( \hat{p}_j \left[ \hat{p}_j^{\dagger}, \hat{O} \right] + \left[ \hat{O}, \hat{p}_j \right] \hat{p}_j^{\dagger} \right) + \gamma_p \sum_{j = 1}^N \left( \hat{p}_j \hat{p}_j^{\dagger} \hat{O} \hat{p}_j \hat{p}_j^{\dagger} - \frac{1}{2} \left\{ \hat{p}_j \hat{p}_j^{\dagger}, \hat{O} \right\} \right) .
    \end{aligned}
\end{equation}
This gives
\begin{equation}
    \begin{aligned}
\left[ \pd{\hat{c}_j} \right]_{\mathrm{sing.}} &= - \frac{\gamma_d}{2} \hat{c}_j - \frac{\gamma_s}{2} \hat{c}_j - \sqrt{\gamma_d} \hat{c}_j^z \hat{\xi}_{\gamma_d,j} + \sqrt{\gamma_s} \hat{r}_j^{\dagger} \hat{\xi}_{\gamma_s,j} - \sqrt{w} \hat{\xi}_{w,j}^{\dagger} \hat{r}_j^{\dagger}, \\
\left[ \pd{\hat{p}_j} \right]_{\mathrm{sing.}} &= - \frac{\gamma_d}{2} \hat{p}_j - \frac{\gamma_s}{2} \hat{p}_j - \frac{w}{2} \hat{p}_j - \frac{\gamma_p}{2} \hat{p}_j + \sqrt{\gamma_d} \hat{r}_j \hat{\xi}_{\gamma_d,j} - \sqrt{\gamma_s} \hat{p}_j^z \hat{\xi}_{\gamma_s,j} + \sqrt{w} \hat{\xi}_{w,j}^{\dagger} \hat{p}_j^z + \sqrt{\gamma_p} \left( \hat{\xi}_{\gamma_p,j}^{\dagger} \hat{p}_j - \hat{p}_j \hat{\xi}_{\gamma_p,j} \right), \\
\left[ \pd{\hat{r}_j} \right]_{\mathrm{sing.}} &= - \frac{w}{2} \hat{r}_j - \frac{\gamma_p}{2} \hat{r}_j - \sqrt{\gamma_d} \hat{\xi}_{\gamma_d,j}^{\dagger} \hat{p}_j - \sqrt{\gamma_s} \hat{c}_j^{\dagger} \hat{\xi}_{\gamma_s,j} + \sqrt{w} \hat{\xi}_{w,j} \hat{c}_j^{\dagger} + \sqrt{\gamma_p} \left( \hat{\xi}_{\gamma_p,j}^{\dagger} \hat{r}_j - \hat{r}_j \hat{\xi}_{\gamma_p,j} \right).
    \end{aligned}
\end{equation}

\paragraph{Total dynamics:}
Now combining the single-particle terms with the collective terms, we find
\begin{equation} \label{MF_HeisenbergLangevin}
    \begin{aligned}
\pd{\hat{c}_-} &= \left[ \pd{\hat{c}_-} \right]_{\mathrm{coll.}} + \left[ \pd{\hat{c}_-} \right]_{\mathrm{sing.}} = \frac{i \Omega}{2} \hat{p}_- + \frac{N W}{2} \hat{p}_- \hat{r}_+ + \frac{N \Gamma_c}{2} \hat{c}_z \hat{c}_- - \frac{\gamma_d}{2} \hat{c}_- - \frac{\gamma_s}{2} \hat{c}_- + \hat{\mathcal{F}}_c, \\
\pd{\hat{p}_-} &= \left[ \pd{\hat{p}_-} \right]_{\mathrm{coll.}} + \left[\pd{\hat{p}_-} \right]_{\mathrm{sing.}} = \frac{i \Omega}{2} \hat{c}_- - \frac{N W}{2} \hat{p}_- \hat{p}_z - \frac{N \Gamma_c}{2} \hat{r}_- \hat{c}_- - \frac{\gamma_d}{2} \hat{p}_- - \frac{\gamma_s}{2} \hat{p}_- - \frac{w}{2} \hat{p}_- - \frac{\gamma_p}{2} \hat{p}_- + \hat{\mathcal{F}}_p, \\
\pd{\hat{r}_-} &= \left[ \pd{\hat{r}_-} \right]_{\mathrm{coll.}} + \left[\pd{\hat{r}_-} \right]_{\mathrm{sing.}} = \frac{i \Omega}{2} \hat{r}_z - \frac{N W}{2} \hat{p}_- \hat{c}_+ + \frac{N \Gamma_c}{2} \hat{c}_+ \hat{p}_- - \frac{w}{2} \hat{r}_- - \frac{\gamma_p}{2} \hat{r}_- + \hat{\mathcal{F}}_r,
    \end{aligned}
\end{equation}
with the Langevin noise
\begin{equation} \label{CoherenceNoise}
    \begin{aligned}
\hat{\mathcal{F}}_c &= - \sqrt{W} \hat{\xi}_W^{\dagger} \hat{r}_+ - \sqrt{\Gamma_c} \hat{c}_z \hat{\xi}_{\Gamma_c} + \frac{1}{N} \sum_{j = 1}^N \left[ - \sqrt{\gamma_d} \hat{c}_j^z \hat{\xi}_{\gamma_d,j} + \sqrt{\gamma_s} \hat{r}_j^{\dagger} \hat{\xi}_{\gamma_s,j} - \sqrt{w} \hat{\xi}_{w,j}^{\dagger} \hat{r}_j^{\dagger} \right], \\
\hat{\mathcal{F}}_p &= \sqrt{W} \hat{\xi}_W^{\dagger} \hat{p}_z + \sqrt{\Gamma_c} \hat{r}_- \hat{\xi}_{\Gamma_c} + \frac{1}{N} \sum_{j = 1}^N \left[ \sqrt{\gamma_d} \hat{r}_j \hat{\xi}_{\gamma_d,j} - \sqrt{\gamma_s} \hat{p}_j^z \hat{\xi}_{\gamma_s,j} + \sqrt{w} \hat{\xi}_{w,j}^{\dagger} \hat{p}_j^z + \sqrt{\gamma_p} \left( \hat{\xi}_{\gamma_p,j}^{\dagger} \hat{p}_j - \hat{p}_j \hat{\xi}_{\gamma_p,j} \right) \right], \\
\hat{\mathcal{F}}_r &= \sqrt{W} \hat{\xi}_W^{\dagger} \hat{c}_+ - \sqrt{\Gamma_c} \hat{\xi}_{\Gamma_c}^{\dagger} \hat{p}_- + \frac{1}{N} \sum_{j = 1}^N \left[ - \sqrt{\gamma_d} \hat{\xi}_{\gamma_d,j}^{\dagger} \hat{p}_j - \sqrt{\gamma_s} \hat{c}_j^{\dagger} \hat{\xi}_{\gamma_s,j} + \sqrt{w} \hat{\xi}_{w,j} \hat{c}_j^{\dagger} + \sqrt{\gamma_p} \left( \hat{\xi}_{\gamma_p,j}^{\dagger} \hat{r}_j - \hat{r}_j \hat{\xi}_{\gamma_p,j} \right) \right].
    \end{aligned}
\end{equation}

\paragraph{Mean-field description:}
We now break the normalized coherences into a mean-field component $b = \expval{\hat{b}_-}$ ($b = c, p, r$) plus fluctuations,
\begin{equation}
    \hat{c}_- = c + \delta \hat{c}_-, \quad \hat{p}_- = p + \delta \hat{p}_-, \quad \hat{r}_- = r + \delta \hat{r}_-.
\end{equation}
The dynamics of the mean-field terms $c$, $p$, and $r$ are given by 
\begin{equation} \label{M}
    \begin{aligned}
\pd{c} &= \frac{i \Omega}{2} p + \frac{N W}{2} p r^* + \frac{N \Gamma_c}{2} c_z c - \frac{\gamma_d}{2} c - \frac{\gamma_s}{2} c , \\
\pd{p} &= \frac{i \Omega}{2} c - \frac{N W}{2} p p_z - \frac{N \Gamma_c}{2} rc - \frac{\gamma_d}{2} p - \frac{\gamma_s}{2} p - \frac{w}{2} p - \frac{\gamma_p}{2} \hat{p}, \\
\pd{r} &= \frac{i \Omega}{2} r_z - \frac{N W}{2} p c^* + \frac{N \Gamma_c}{2}c^* p - \frac{w}{2} r - \frac{\gamma_p}{2} r.
    \end{aligned}
\end{equation}
It can be fully described by a single, mean-field $3\times 3$ density matrix $\hat{\rho}_{\mathrm{mf}}$ with $b = \mathrm{Tr}_{\mathrm{mf}} \left[ \hat{b} \hat{\rho}_{\mathrm{mf}} \right]$, where $\hat{b}$ can be $\hat{c} = \hat{\sigma}_{du}$, $\hat{p} = \hat{\sigma}_{su}$, $\hat{r}=\hat{\sigma}_{sd}$, $\hat{r}_z = \hat{\sigma}_{dd}-\hat{\sigma}_{ss}$. 
We can then extract from $\partial b/(\partial t) = \mathrm{Tr}_{\mathrm{mf}}\left[ \hat{b}\hat{\mathcal{L}}_{\mathrm{mf}} \hat{\rho}_{\mathrm{mf}} \right]$ the dynamics of the mean-field density matrix,
\begin{equation} \label{MF_MasterEq}
    \pd{\hat{\rho}_{\mathrm{mf}}} = \hat{\mathcal{L}}_{\mathrm{mf}} \hat{\rho}_{\mathrm{mf}} \equiv - \frac{i}{\hbar} \left[ \hat{H}_{\mathrm{mf}}, \hat{\rho}_{\mathrm{mf}} \right] + \hat{\mathcal{D}} \left[ \sqrt{\gamma_d} \hat{c} \right] \hat{\rho}_{\mathrm{mf}} + \hat{\mathcal{D}} \left[ \sqrt{\gamma_s} \hat{p} \right] \hat{\rho}_{\mathrm{mf}} + \hat{\mathcal{D}} \left[ \sqrt{w} \hat{p}^{\dagger} \right] \hat{\rho}_{\mathrm{mf}} + \hat{\mathcal{D}} \left[ \sqrt{\gamma_p} \hat{p} \hat{p}^{\dagger} \right] \hat{\rho}_{\mathrm{mf}},
\end{equation}
with the mean-field Hamiltonian
\begin{equation} \label{H_mf}
    \hat{H}_{\mathrm{mf}} = \frac{\hbar \Omega}{2} \left( \hat{r} + \hat{r}^{\dagger} \right) + \frac{i \hbar N \Gamma_c}{2} \left( c^* \hat{c} - \hat{c}^{\dagger} c \right) + \frac{i \hbar N W}{2} \left( p \hat{p}^{\dagger} - \hat{p} p^* \right).
\end{equation}
This is a non-linear master equation since the parameters $c$ and $p$ themselves depend on the density matrix $\hat{\rho}_{\mathrm{mf}}$. 
We can self-consistently solve for $c$ and $p$ for a given set of parameters, and then use these values in $\hat{H}_{\mathrm{mf}}$ in order to find the kernel of the mean-field Liouvillian $\hat{\mathcal{L}}_{\mathrm{mf}} \hat{\rho}_{\mathrm{ss}} = 0$.
Finding the quantities self-consistently then allows us to calculate mean-field predictions of all observables.  

\begin{figure}
    \centerline{\includegraphics[width=\linewidth]{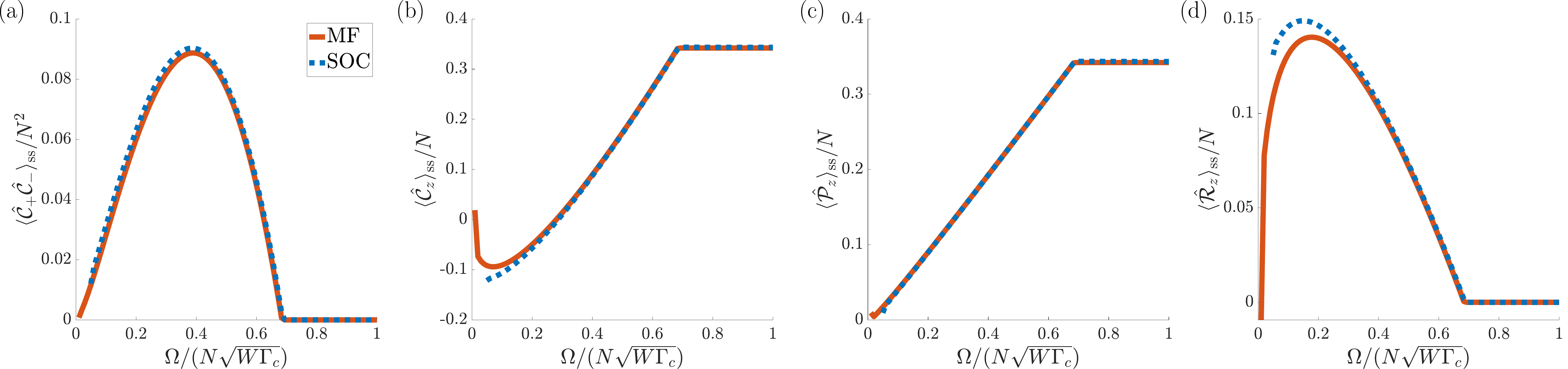}}
    \caption{Comparison between mean-field (MF; solid orange lines) and second-order cumulants (SOC; dashed blue lines) for the Ba parameters of Table~\ref{tb:BaParams} with $N = 10^6$.}
    \label{MF_SOC_Fig}
\end{figure}
In order to check the validity of our mean-field model, we compare steady-state expectation values calculated using mean-field to those calculated using a second-order cumulant approximation. 
We compare the mean-field to a cumulant approximation rather than the full master equation as our mean-field model requires a non-negligible $\hat{\mathcal{L}}_{\mathrm{sp}}$ to reach steady-state, as shown in Eq.~\eqref{MF_MasterEq}, which will remove the state from the polynomial-scaled subspace used in Sec.~\ref{Sec:U1}~\cite{Reilly3} for our exact simulations. 
We use the second-order cumulant method of Ref.~\cite{Kubo} where third-order cumulants are assumed to be zero, which allows one to factorize third-order moments as
\begin{equation}
    \expval{\hat{O}_1 \hat{O}_2 \hat{O}_3} \approx \expval{\hat{O}_1 \hat{O}_2} \expval{\hat{O}_3} + \expval{\hat{O}_1 \hat{O}_3} \expval{\hat{O}_2} + \expval{\hat{O}_1} \expval{\hat{O}_2 \hat{O}_3} - 2 \expval{\hat{O}_1} \expval{\hat{O}_2} \expval{\hat{O}_3}. 
\end{equation}
Using the permutation symmetry of the system, $\expval{\hat{\sigma}_{ij}^{(k)}} = \expval{\hat{\sigma}_{ij}^{(1)}}$ and $\expval{\hat{\sigma}_{ij}^{(k)} \hat{\sigma}_{lm}^{(n)}} = \expval{\hat{\sigma}_{ij}^{(1)} \hat{\sigma}_{lm}^{(2)}} = \expval{\hat{\sigma}_{lm}^{(1)} \hat{\sigma}_{ij}^{(2)}}$, one finds a set of $33$ non-linear coupled equations for first- and second-order moments consisting of the operators $\hat{O}_1, \hat{O}_2 \in \left\{ \hat{\sigma}_{uu}^{(j)}, \hat{\sigma}_{dd}^{(j)}, \hat{\sigma}_{ss}^{(j)}, \hat{\sigma}_{ud}^{(j)}, \hat{\sigma}_{us}^{(j)}, \hat{\sigma}_{ds}^{(j)} \right\}$ for $j \in \{ 1,2 \}$. 
By further using the additional $\mathrm{U} (1)$ symmetry discussed in Sec.~\ref{Sec:U1}, this reduces the set of coupled non-linear equations to $18$ (for a suitable initial state such as all atoms in $\ket{d}$) which can then be numerically integrated to steady-state. 
We have checked that the second-order cumulant steady-state results agree well with the full master equation steady-state results for the observables of interest in the case of $\hat{\mathcal{L}}_{\mathrm{sp}} = 0$.
We display the comparison between mean-field and cumulants in Fig.~\ref{MF_SOC_Fig} where we show the $x$-cavity intensity and the respective inversions of the three transitions $\mathcal{C}$, $\mathcal{P}$, and $\mathcal{R}$. 
We find remarkable agreement between the mean-field and the second-order cumulant models throughout the parameter regime of interest, namely the region of peak intensity.
The main difference between the mean-field results occurs at low drive strengths $\Omega \lesssim 0.2 N \sqrt{W \Gamma_c}$. 
This corresponds to $\Omega \lesssim 178 \, \mathrm{Hz}$ for the barium parameters of Table~\ref{tb:BaParams} with $N = 10^6$.

\subsection{Superradiant lasing threshold}
We now attempt to find the threshold in our parameters in order to achieve steady-state superradiant lasing. 
In this section, we treat the single-particle dissipation perturbatively.
To find the lasing thresholds, we will assume that the system is in a non-lasing steady-state and then check whether the non-lasing state is stable against fluctuations. 
Such fluctuations are introduced by the initial state and also by noise coming from dissipation. 
However, only the response of the atoms is relevant for the determination of their stability while details on the fluctuations would be only important for describing the detailed dynamics. 
The equations of motion for the fluctuations, $\delta \hat{o} = \hat{o} - \expval{\hat{o}}$, in the absence of $\hat{\mathcal{L}}_{\mathrm{sp}}$, are given by
\begin{equation} \label{Noise_EoM}
    \begin{aligned}
\pd{(\delta \hat{c}_-)} &= \frac{i \Omega}{2} \delta \hat{p}_- + \frac{N W}{2} \left( \expval{\hat{p}_-} \delta \hat{r}_+ + \delta \hat{p}_- \expval{\hat{r}_+} \right) + \frac{N \Gamma_c}{2} \expval{\hat{c}_z} \delta \hat{c}_-, \\
\pd{(\delta \hat{p}_-)} &= \frac{i \Omega}{2} \delta \hat{c}_- - \frac{N W}{2} \delta \hat{p}_- \expval{\hat{p}_z} - \frac{N \Gamma_c}{2} \left( \expval{\hat{r}_-} \delta \hat{c}_- + \delta \hat{r}_- \expval{\hat{c}_-} \right), \\
\pd{(\delta \hat{r}_-)} &= \frac{i \Omega}{2} \expval{\hat{r}_z} - \frac{N W}{2} \left( \expval{\hat{p}_-} \delta \hat{c}_+ + \delta \hat{p}_- \expval{\hat{c}_+} \right) + \frac{N \Gamma_c}{2} \left( \expval{\hat{c}_+} \delta \hat{p}_- + \delta \hat{c}_+ \expval{\hat{p}_-} \right).
    \end{aligned}
\end{equation}
Here, we have dropped the noise contribution because it is irrelevant for determining the threshold. 

\subsubsection{All atoms in the ground states}
Fixing $\Omega$ at a suitably large value, the regime $W \ll \Gamma_c, \Omega / N$ leads to the system essentially just performing damped Rabi flopping.
Here, we have an incoherent mixture of $\hat{\rho}_{\mathrm{mf}} = (\op{d}{d} + \op{s}{s}) / 2$ which gives $\expval{\hat{r}_z}_{\mathrm{ss}} \approx 0$ while $\expval{\hat{c}_z}_{\mathrm{ss}} \approx \expval{\hat{p}_z}_{\mathrm{ss}} \approx - 1 / 2$.
Meanwhile, the coherences all decay to zero, $\expval{\hat{c}_-}_{\mathrm{ss}} \approx \expval{\hat{p}_-}_{\mathrm{ss}} \approx \expval{\hat{r}_-}_{\mathrm{ss}} \approx 0$. 
Plugging these steady-state values into Eq.~\eqref{Noise_EoM}, we find
\begin{equation}
    \begin{aligned}
\pd{(\delta \hat{c}_-)} &= \frac{i \Omega}{2} \delta \hat{p}_- - \frac{N \Gamma_c}{4} \delta \hat{c}_-, \\
\pd{(\delta \hat{p}_-)} &= \frac{i \Omega}{2} \delta \hat{c}_- + \frac{N W}{4} \delta \hat{p}_-,
    \end{aligned}
\end{equation}
which gives the two-dimensional coupling matrix,
\begin{equation}
    \pd{} \binom{\delta \hat{c}_-}{\delta \hat{p}_-} = \frac{1}{4} \begin{pmatrix}
        - N \Gamma_c & 2 i \Omega \\
        2 i \Omega & N W
    \end{pmatrix} \binom{\delta \hat{c}_-}{\delta \hat{p}_-} \equiv \bm{A} \binom{\delta \hat{c}_-}{\delta \hat{p}_-}.
\end{equation}
We can first find the eigenvalues $\lambda_{\pm}$ of the coupling matrix $\bm{A}$,
\begin{equation}
    \lambda_{\pm} = \frac{N (W - \Gamma_c)}{8} \pm \frac{1}{8} \sqrt{(N \Gamma_c + N W)^2 - 16 \Omega^2},
\end{equation}
such that for appropriate linear combinations of the fluctuations, i.e. the eigenstates of $\bm{A}$, we obtain
\begin{equation}
    \pd{} \vec{v} = 
    \begin{pmatrix}
        \lambda_+ & 0 \\
        0 & \lambda_-
    \end{pmatrix} \vec{v}.
\end{equation}
Thus, for $\vec{v} = \binom{v_+}{v_-}$, we find the solutions
\begin{equation}
    v_+ = e^{\lambda_+ t}, \quad v_- = e^{\lambda_- t}.
\end{equation}
Therefore, the stability of the solution breaks down when $\Re[\lambda_+] > 0$ or $\Re[\lambda_-] > 0$. 
One can immediately see that $\Re[\lambda_+] > 0$ when $W > \Gamma_c$, and so the first stability condition is simply
\begin{equation}
    W < \Gamma_c. 
\end{equation}
Meanwhile, for the second condition, we find that $\lambda_+ = 0$ is equivalent to
\begin{equation}
    W^2 - 2 W \Gamma_c + \Gamma_c^2 = \Gamma_c^2 + 2 W \Gamma_c + W^2 - \frac{16 \Omega^2}{N^2}.
\end{equation}
This reduces to our second stability condition to
\begin{equation} \label{OmegaIneq}
    W < \frac{4 \Omega^2}{N^2 \Gamma_c}.
\end{equation}

\subsubsection{All atoms in the excited state}
In the opposite limit, $W \gg \Omega / N > \Gamma_c$, the atoms are pumped into the $\ket{u}$ state and thus we have to determine the stability of $\hat{\rho}_{\mathrm{mf}} = \op{u}{u}$. 
This state gives $\expval{\hat{r}_z}_{\mathrm{ss}} \approx 0$, $\expval{\hat{c}_z}_{\mathrm{ss}} \approx \expval{\hat{p}_z}_{\mathrm{ss}} \approx 1$, and again $\expval{\hat{c}_-}_{\mathrm{ss}} \approx \expval{\hat{p}_-}_{\mathrm{ss}} \approx \expval{\hat{r}_-}_{\mathrm{ss}} \approx 0$.
Plugging these steady-state values into Eq.~\eqref{Noise_EoM}, we find
\begin{equation}
    \begin{aligned}
\pd{(\delta \hat{c}_-)} &= \frac{i \Omega}{2} \delta \hat{p}_- + \frac{N \Gamma_c}{2} \delta \hat{c}_-, \\
\pd{(\delta \hat{p}_-)} &= \frac{i \Omega}{2} \delta \hat{c}_- - \frac{N W}{2} \delta \hat{p}_-,
    \end{aligned}
\end{equation}
which gives the two-dimensional coupling matrix
\begin{equation}
    \pd{} \binom{\delta \hat{c}_-}{\delta \hat{p}_-} = \frac{1}{2} \begin{pmatrix}
        N \Gamma_c & i \Omega \\
        i \Omega & - N W
    \end{pmatrix} \binom{\delta \hat{c}_-}{\delta \hat{p}_-} \equiv \bm{B} \binom{\delta \hat{c}_-}{\delta \hat{p}_-}.
\end{equation}
The eigenvalues of $\bm{B}$ are given by
\begin{equation}
    \lambda_{\pm} = \frac{N (\Gamma_c - W)}{4} \pm \frac{1}{4} \sqrt{(N \Gamma_c + N W)^2 - 4 \Omega^2}.
\end{equation}
The first condition for stability is then
\begin{equation}
    W > \Gamma_c,
\end{equation}
while the solution $\lambda_+ = 0$, which is equivalent to 
\begin{equation}
    \Gamma_c^2 - 2 W \Gamma_c + W^2 = \Gamma_c^2 + 2 W \Gamma_c + W^2 - 4 \frac{\Omega^2}{N^2},
\end{equation}
again gives the second stability condition,
\begin{equation}
    W < \frac{\Omega^2}{N^2 \Gamma_c}. 
\end{equation}

\subsubsection{Cases}
Combining our four conditions for stability together, we get three cases for the thresholds:
\begin{enumerate}
    \item Case $\Omega < \frac{N \Gamma_c}{2}$: We expect lasing when $W > \frac{4 \Omega^2}{N^2 \Gamma_c}$ and no lasing when $W < \frac{4 \Omega^2}{N^2 \Gamma_c}$, and so the threshold is $\Omega = \frac{N \sqrt{W \Gamma_c}}{2}$. 

    \item Case $\frac{N \Gamma_c}{2} < \Omega < N \Gamma_c$: We expect lasing when $W > \Gamma_c$ and no lasing when $W < \Gamma_c$, and so the threshold is $W = \Gamma_c$. 

    \item Case $N \Gamma_c < \Omega$: We expect lasing when $W > \frac{\Omega^2}{N^2 \Gamma_c}$ and no lasing when $W < \frac{\Omega^2}{N^2 \Gamma_c}$, and so the threshold is $\Omega = N \sqrt{W \Gamma_c}$.
\end{enumerate}

\subsection{Linewidth} \label{Sec:MF_linewidth}
We now calculate the linewidth of the output light. 
The general idea is to use a phase diffusion argument which allows us to determine the coherence time of the mean-field quantities $c$, $p$, and $r$. 
The mean-field solution breaks the $\mathrm{U}(1)$ symmetry and the first step is to choose an appropriate axis for the mean-field quantities. We choose $c \in \mathbb{R}$. 
In order to be able to solve the mean-field equations of motions in Eq.~\eqref{M}, we thus need to choose the other two coherences imaginary $p, r \in i \mathbb{R}$.
We are interested in the phase diffusion of the $x$-cavity, so we write
\begin{equation}
    \hat{c}_- = \left( \abs{c} + \abs{\delta \hat{c}_-} \right) e^{i \varphi_c + i \delta \hat{\varphi}_c} \approx \left( \abs{c} + \abs{\delta \hat{c}_-} \right) e^{i \varphi_c} \left( 1 + i \delta \hat{\varphi}_c \right) \approx \abs{c} e^{i \varphi_c} + \abs{\delta \hat{c}_-} e^{i \varphi_c} + i \delta \hat{\varphi}_c \abs{c} e^{i \varphi_c},
\end{equation}
and so we find 
\begin{equation}
    \delta \hat{\varphi}_c \approx \Im \left[ \frac{\delta \hat{c}_-}{c} \right],
\end{equation}
and accordingly 
\begin{equation}
    \delta \hat{\varphi}_p \approx \Im \left[ \frac{\delta \hat{p}_-}{p} \right],\quad
    \delta \hat{\varphi}_r \approx \Im \left[ \frac{\delta \hat{r}_-}{r} \right].
\end{equation}
From Eq.~\eqref{MF_HeisenbergLangevin} we can then derive coupled equations for the angles
\begin{equation}
    \begin{aligned}
\pd{} \delta \hat{\varphi}_c &= \frac{i \Omega p}{2 c} \delta \hat{\varphi}_p + \frac{N W p r^*}{2 c} \left( \delta \hat{\varphi}_p - \delta \hat{\varphi}_r \right) + \frac{N \Gamma_c c_z}{2} \delta \hat{\varphi}_c - \frac{\gamma_d}{2} \delta \hat{\varphi}_c - \frac{\gamma_s}{2} \delta \hat{\varphi}_c + \Im \left[ \frac{\hat{\mathcal{F}}_c}{c} \right], \\
\pd{} \delta \hat{\varphi}_p &= \frac{i \Omega c}{2 p} \delta \hat{\varphi}_c - \frac{N W p_z}{2} \delta \hat{\varphi}_p - \frac{N \Gamma_c r c}{2 p} \left( \delta \hat{\varphi}_r + \delta \hat{\varphi}_c \right) - \frac{\gamma_d}{2} \delta \hat{\varphi}_p - \frac{\gamma_s}{2} \delta \hat{\varphi}_p - \frac{w}{2} \delta \hat{\varphi}_p - \frac{\gamma_p}{2} \delta \hat{\varphi}_p + \Im \left[ \frac{\hat{\mathcal{F}}_p}{p} \right], \\
\pd{} \delta \hat{\varphi}_r &= \frac{N W p c^*}{2 r} \left[ \delta \hat{\varphi}_c - \delta \hat{\varphi}_p \right] + \frac{N \Gamma_c c^* p}{2 r} \left[ \delta \hat{\varphi}_p - \delta \hat{\varphi}_c \right] - \frac{w}{2} \delta \hat{\varphi}_r - \frac{\gamma_p}{2} \delta \hat{\varphi}_r + \Im \left[ \frac{\hat{\mathcal{F}}_r}{r} \right],
    \end{aligned}
\end{equation}
where we have linearized the equations and explicitly used $c \in \mathbb{R}$ and $p, r \in i \mathbb{R}$.
Vectorizing, $\vec{\bm{\varphi}} = (\delta \hat{\varphi}_c, \delta \hat{\varphi}_p, \delta \hat{\varphi}_r)^T$, we can write this as
\begin{equation} \label{phi_EoM}
    \pd{\vec{\bm{\varphi}}} = \bm{M} \vec{\bm{\varphi}} + \vec{\bm{\mathcal{N}}},
\end{equation}
with the coupling matrix
\begin{equation}
    \bm{M} \equiv \frac{1}{2}
\begin{pmatrix}
    N \Gamma_c c_z - \gamma_d - \gamma_s & i \Omega \frac{p}{c} + N W \frac{p r^*}{c} & - N W \frac{p r^*}{c} \\
    i \Omega \frac{c}{p} - N \Gamma_c \frac{r c}{p} & - N W p_z - \gamma_d - \gamma_s - w - \gamma_p & - N \Gamma_c \frac{r c}{p} \\
    \frac{p c^*}{r} [N W - N \Gamma_c] & \frac{p c^*}{r} [N \Gamma_c - N W] & - w - \gamma_p
\end{pmatrix},
\end{equation}
and the noise vector
\begin{equation}
    \vec{\bm{\mathcal{N}}} \equiv 
\begin{pmatrix}
    \Im \left[ \frac{\hat{\mathcal{F}}_c}{c} \right] \\[1.5 mm]
    \Im \left[ \frac{\hat{\mathcal{F}}_p}{p} \right] \\[1.5 mm]
    \Im \left[ \frac{\hat{\mathcal{F}}_r}{r} \right]
\end{pmatrix}.
\end{equation}
From Eq.~\eqref{CoherenceNoise}, we see that the Langevin noise from single-particle decoherence can be ignored in the limit $N \rightarrow \infty$. 
We therefore approximate 
\begin{equation}
    \vec{\bm{\mathcal{N}}} \approx - \frac{i}{2}
\begin{pmatrix}
    - \sqrt{W} \frac{r^*}{c} \hat{\xi}_W^{\dagger} - \sqrt{\Gamma_c} \frac{c_z}{c} \hat{\xi}_{\Gamma_c} + \sqrt{W} \frac{r}{c^*} \hat{\xi}_W + \sqrt{\Gamma_c} \frac{c_z}{c^*} \hat{\xi}_{\Gamma_c}^{\dagger} \\[1.5 mm]
    \sqrt{W} \frac{p_z}{p} \hat{\xi}_W^{\dagger} + \sqrt{\Gamma_c} \frac{r}{p} \hat{\xi}_{\Gamma_c} - \sqrt{W} \frac{p_z}{p^*} \hat{\xi}_W - \sqrt{\Gamma_c} \frac{r^*}{p^*} \hat{\xi}_{\Gamma_c}^{\dagger} \\[1.5 mm]
    \sqrt{W} \frac{c^*}{r} \hat{\xi}_W^{\dagger} - \sqrt{\Gamma_c} \frac{p}{r} \hat{\xi}_{\Gamma_c}^{\dagger} - \sqrt{W} \frac{c}{r^*} \hat{\xi}_W + \sqrt{\Gamma_c} \frac{p^*}{r^*} \hat{\xi}_{\Gamma_c} 
\end{pmatrix} = - \frac{i}{2}
\begin{pmatrix}
    \sqrt{W} \frac{r}{c} \left( \hat{\xi}_W^{\dagger} + \hat{\xi}_W \right) + \sqrt{\Gamma_c} \frac{c_z}{c} \left( \hat{\xi}_{\Gamma_c}^{\dagger} - \hat{\xi}_{\Gamma_c} \right) \\[1.5 mm]
    \sqrt{W} \frac{p_z}{p} \left( \hat{\xi}_W^{\dagger} + \hat{\xi}_W \right) + \sqrt{\Gamma_c} \frac{r}{p} \left( \hat{\xi}_{\Gamma_c} - \hat{\xi}_{\Gamma_c}^{\dagger} \right) \\[1.5 mm]
    \sqrt{W} \frac{c}{r} \left( \hat{\xi}_W^{\dagger} + \hat{\xi}_W \right) + \sqrt{\Gamma_c} \frac{p}{r} \left( \hat{\xi}_{\Gamma_c} - \hat{\xi}_{\Gamma_c}^{\dagger} \right)
\end{pmatrix},
\end{equation}
where we have used $c, c_z, p_z \in \mathbb{R}$ and $p, r \in i \mathbb{R}$. 
The corresponding diffusion matrix, $\expval{\vec{\bm{\mathcal{N}}} (t) \vec{\bm{\mathcal{N}}}^T (t')} = 2 \bm{D} \delta (t - t')$, thus reads
\begin{equation}
    \bm{D} = - \frac{1}{8}
\begin{pmatrix}
    W \frac{r^2}{c^2} - \Gamma_c \frac{c_z^2}{c^2} & W \frac{r p_z}{c p} + \Gamma_c \frac{c_z r}{c p} & W + \Gamma_c \frac{c_z p}{c r} \\
    W \frac{p_z r}{p c} + \Gamma_c \frac{r c_z}{p c} & W \frac{p_z^2}{p^2} - \Gamma_c \frac{r^2}{p^2} & W \frac{p_z c}{p r} - \Gamma_c \\
    W + \Gamma_c \frac{p c_z}{r c} & W \frac{c p_z}{r p} - \Gamma_c & W \frac{c^2}{r^2} - \Gamma_c \frac{p^2}{r^2}
\end{pmatrix}.
\end{equation}

To calculate the linewidth, we formally integrate Eq.~\eqref{phi_EoM} and obtain
\begin{equation}
    \vec{\bm{\varphi}} (t) = e^{\bm{M} t} \vec{\bm{\varphi}} (0) + \int_0^t d \tau e^{\bm{M} (t - \tau)} \vec{\bm{\mathcal{N}}} (\tau).
\end{equation}
Using the eigen-decomposition of the coupling matrix, $\bm{M} = \bm{V} \bm{\Lambda} \bm{V}^{-1}$, we have
\begin{equation}
    \vec{\bm{\varphi}} (t) = \bm{V} e^{\bm{\Lambda} t} \bm{V}^{-1} \vec{\bm{\varphi}} (0) + \int_0^t d \tau \bm{V} e^{\bm{\Lambda} (t - \tau)} \bm{V}^{-1} \vec{\bm{\mathcal{N}}} (\tau).
\end{equation}
Here, $\bm{\Lambda}=\mathrm{diag}(\lambda_0,\lambda_1,\lambda_2)$ and ${\bf V}=(\vec{\bm{v}}_0,\vec{\bm{v}}_1,\vec{\bm{v}}_2)$ are the matrix of eigenvalues $\lambda_i$ and the corresponding eigenvectors~$\vec{\bm{v}}_i$. 
If the lasing regime is stable, we find one zero eigenvalue $\lambda_0=0$ and two negative eigenvalues $\lambda_1,\lambda_2<0$.
The appearance of a zero eigenvalue is a direct consequence of the $\mathrm{U}(1)$ symmetry which implies the existence of a Goldstone mode~\cite{Goldstone}. 
Since $\bm{\Lambda}$ is a diagonal matrix with one zero ($\lambda_0$) and two negative numbers ($\lambda_1$ and $\lambda_2$), the first term can be ignored since it is either decaying or a constant offset.
We can therefore calculate
\begin{equation}
    \Delta \varphi_c^2 = \expval{\vec{\bm{e}}_1^T \vec{\bm{\varphi}} (t) \vec{\bm{\varphi}}^T (t) \vec{\bm{e}}_1} = 2 \int_0^t d \tau \int_0^t d \tau' \vec{\bm{e}}_1^T \bm{V} e^{\bm{\Lambda} (t - \tau)} \bm{V}^{-1} \bm{D} \delta(\tau - \tau') \left( \bm{V}^{\dagger} \right)^{-1} e^{\bm{\Lambda}^{\dagger} (t - \tau')} \bm{V}^{\dagger} \vec{\bm{e}}_1,
\end{equation}
where $\vec{\bm{e}}_1 = (1, 0, 0)^T$ is used to get the $c$ component.
Since the population is dominated by the $\vec{\bm{v}}_0$ eigenstate (corresponding to the Goldstone mode), the linewidth of the $x$-cavity is approximately given by
\begin{equation}
    \Delta \nu = \Delta \varphi_c^2 / t \approx 2 \abs{\vec{\bm{e}}_1^T \vec{\bm{v}}_0}^2 \vec{\bm{e}}_1^T \bm{V}^{-1} \bm{D} \left( \bm{V}^{\dagger} \right)^{-1} \vec{\bm{e}}_1.
\end{equation}
This allows us to calculate the linewidth and only requires the pre-calculation of the mean-field quantities $c$, $p$, $r$, $c_z$, $p_z$, and $r_z$.
This approach is used to compute the linewidth visible in Fig.~3(h) of the main text. 

\subsection{Cavity pulling}
We now examine how changes in the cavity length, such as thermal fluctuations, change the frequency of the light output from the cavity. 
Here, we focus on fluctuations of the $x$-oriented cavity. 
To do this, we note that changes in cavity length will in turn change the cavity frequency $\omega_x$~\cite{Steck}, and so we now allow $\Delta_x$ to become non-zero. 
The system's Hamiltonian in Eq.~\eqref{MasterEq_nzDeltas} therefore becomes
\begin{equation}
    \hat{H} = \hbar \Omega \hat{\mathcal{R}}_x - \hbar \chi_x \hat{\mathcal{C}}_+ \hat{\mathcal{C}}_-,
\end{equation}
which we note does not break the $\mathrm{U} (1)$ symmetry discussed in Section~\ref{Sec:U1}.
This modifies the collective equations of motion of the coherences Eq.~\eqref{MF_Coherences} by 
\begin{equation}
    \begin{aligned}
\pd{\hat{\mathcal{C}}_-} &= - 2 i \chi_x \hat{\mathcal{C}}_z \hat{\mathcal{C}}_- + \frac{i \Omega}{2} \hat{\mathcal{P}}_- + \frac{W}{2} \hat{\mathcal{P}}_- \hat{\mathcal{R}}_+ + \Gamma_c \hat{\mathcal{C}}_z \hat{\mathcal{C}}_- - \sqrt{W} \hat{\xi}_W^{\dagger} \hat{\mathcal{R}}_+ - 2 \sqrt{\Gamma_c} \hat{\mathcal{C}}_z \hat{\xi}_{\Gamma_c}, \\
\pd{\hat{\mathcal{P}}_-} &= i \chi_x \hat{\mathcal{R}}_- \hat{\mathcal{C}}_- + \frac{i \Omega}{2} \hat{\mathcal{C}}_- - W \hat{\mathcal{P}}_- \hat{\mathcal{P}}_z - \frac{\Gamma_c}{2} \hat{\mathcal{R}}_- \hat{\mathcal{C}}_- + 2 \sqrt{W} \hat{\xi}_W^{\dagger} \hat{\mathcal{P}}_z + \sqrt{\Gamma_c} \hat{\mathcal{R}}_- \hat{\xi}_{\Gamma_c}, \\
\pd{\hat{\mathcal{R}}_-} &= i \chi_x \hat{\mathcal{C}}_+ \hat{\mathcal{P}}_- + i \Omega \hat{\mathcal{R}}_z - \frac{W}{2} \hat{\mathcal{P}}_- \hat{\mathcal{C}}_+ + \frac{\Gamma_c}{2} \hat{\mathcal{C}}_+ \hat{\mathcal{P}}_- + \sqrt{W} \hat{\xi}_W^{\dagger} \hat{\mathcal{C}}_+ - \sqrt{\Gamma_c} \hat{\xi}_{\Gamma_c}^{\dagger} \hat{\mathcal{P}}_-.
    \end{aligned}
\end{equation}
Normalizing by $N$ and adding in the effects of $\hat{\mathcal{L}}_{\mathrm{sp}}$, we find
\begin{equation}
    \begin{aligned}
\pd{\hat{c}_-} &= - i N \chi_x \hat{c}_z \hat{c}_- + \frac{i \Omega}{2} \hat{p}_- + \frac{N W}{2} \hat{p}_- \hat{r}_+ + \frac{N \Gamma_c}{2} \hat{c}_z \hat{c}_- - \frac{\gamma_d}{2} \hat{c}_- - \frac{\gamma_s}{2} \hat{c}_- + \hat{\mathcal{F}}_c, \\
\pd{\hat{p}_-} &= i N \chi_x \hat{r}_- \hat{c}_- + \frac{i \Omega}{2} \hat{c}_- - \frac{N W}{2} \hat{p}_- \hat{p}_z - \frac{N \Gamma_c}{2} \hat{r}_- \hat{c}_- - \frac{\gamma_d}{2} \hat{p}_- - \frac{\gamma_s}{2} \hat{p}_- - \frac{w}{2} \hat{p}_- - \frac{\gamma_p}{2} \hat{p}_- + \hat{\mathcal{F}}_p, \\
\pd{\hat{r}_-} &= i N \chi_x \hat{c}_+ \hat{p}_- + \frac{i \Omega}{2} \hat{r}_z - \frac{N W}{2} \hat{p}_- \hat{c}_+ + \frac{N \Gamma_c}{2} \hat{c}_+ \hat{p}_- - \frac{w}{2} \hat{r}_- - \frac{\gamma_p}{2} \hat{r}_- + \hat{\mathcal{F}}_r,
    \end{aligned}
\end{equation}
with the same Langevin noise $\hat{\mathcal{F}}_i$ as Eq.~\eqref{CoherenceNoise}. 
This modifies the mean-field Hamiltonian Eq.~\eqref{H_mf} into
\begin{equation}
    \hat{H}_{\mathrm{mf}} = \frac{\hbar \Omega}{2} \left( \hat{r} + \hat{r}^{\dagger} \right) - \hbar N \chi_x \left( c^* \hat{c} + \hat{c}^{\dagger} c \right) + \frac{i \hbar N \Gamma_c}{2} \left( c^* \hat{c} - \hat{c}^{\dagger} c \right) + \frac{i \hbar N W}{2} \left( p \hat{p}^{\dagger} - \hat{p} p^* \right).
\end{equation}
To extract the cavity frequency $\omega$ at the oscillatory steady-state, we assume $\expval{\hat{c}_-} \approx \abs{c} e^{i (\omega - \tilde{\omega}_d) t}$, where we noted that we are in a rotating frame where the $x$-cavity rotates with $\tilde{\omega}_d$, which gives
\begin{equation}
    \pd{\expval{\hat{c}_-}} = \pd{\abs{c}} e^{i (\omega - \tilde{\omega}_d) t} + i (\omega - \tilde{\omega}_d) \expval{\hat{c}_-} = - i N \chi_x \expval{\hat{c}_z} \expval{\hat{c}_-} + \frac{i \Omega}{2} \expval{\hat{p}_-} + \frac{N W}{2} \expval{\hat{p}_-} \expval{\hat{r}_+} + \left( \frac{N \Gamma_c}{2} \expval{\hat{c}_z} - \frac{\gamma_d}{2} - \frac{\gamma_s}{2} \right) \expval{\hat{c}_-}.
\end{equation}
We divide by $\expval{\hat{c}_-}$, take the imaginary part, and consider the oscillatory steady-state $\rpd \abs{c} = 0$ to find
\begin{equation} \label{omegaEq}
    \omega = \tilde{\omega}_d - N \chi_x \expval{\hat{c}_z}_{\mathrm{ss}} + \frac{\Omega}{2} \Re \left[ \frac{\expval{\hat{p}_-}_{\mathrm{ss}}}{\expval{\hat{c}_-}_{\mathrm{ss}}} \right] + \frac{N W}{2} \Im \left[ \frac{\expval{\hat{p}_-}_{\mathrm{ss}} \expval{\hat{r}_+}_{\mathrm{ss}}}{\expval{\hat{c}_-}_{\mathrm{ss}}} \right],
\end{equation}
where we have used $\Im[\expval{\hat{c}_z}] = 0$. 
This therefore points to why $\Omega^{\wp}_0$ does not necessarily align with $\Omega^z_0$ in Fig.~3(f) of the main text as there can be additional shifts from the dipoles. 
When $\chi_x = 0$, we find $\omega = \tilde{\omega}_d$ as expected. 

While what is presented above is sufficient to time evolve the system to the oscillatory steady-state, a non-zero laser frequency no longer allows us to simply find the null space of the mean-field Liouvillian; instead, we have~\cite{Jager2} $\hat{\mathcal{L}}_{\mathrm{mf}} \hat{\rho}_{\mathrm{ss}} = i (\omega - \tilde{\omega}_d) [\hat{c}^{\dagger} \hat{c}, \hat{\rho}_{\mathrm{ss}}]$.
To simulate the system's steady-state with the kernel method, we must now move into an interaction picture where $\ket{u}$ rotates with this frequency $\omega - \tilde{\omega}_d$
such that the mean-field Hamiltonian becomes
\begin{equation}
    \tilde{\hat{H}}_{\mathrm{mf}} = \hbar (\omega - \tilde{\omega}_d) \hat{c}^{\dagger} \hat{c} + \frac{\hbar \Omega}{2} \left( \hat{r} + \hat{r}^{\dagger} \right) - \hbar N \chi_x \left( c^* \hat{c} + \hat{c}^{\dagger} c \right) + \frac{i \hbar N \Gamma_c}{2} \left( c^* \hat{c} - \hat{c}^{\dagger} c \right) + \frac{i \hbar N W}{2} \left( p \hat{p}^{\dagger} - \hat{p} p^* \right).
\end{equation}
The master equation for the rotating density matrix $\tilde{\hat{\rho}}_{\mathrm{mf}} = \exp[i (\tilde{\omega}_d - \omega) \hat{c}^{\dagger} \hat{c} t] \hat{\rho}_{\mathrm{mf}} \exp[i (\omega - \tilde{\omega}_d) \hat{c}^{\dagger} \hat{c} t]$ is then given by
\begin{equation} \label{MF_MasterEq_rot}
    \pd{\tilde{\hat{\rho}}_{\mathrm{mf}}} = \tilde{\hat{\mathcal{L}}}_{\mathrm{mf}} \tilde{\hat{\rho}}_{\mathrm{mf}} \equiv - \frac{i}{\hbar} \left[ \tilde{\hat{H}}_{\mathrm{mf}}, \tilde{\hat{\rho}}_{\mathrm{mf}} \right] + \hat{\mathcal{D}} \left[ \sqrt{\gamma_d} \hat{c} \right] \tilde{\hat{\rho}}_{\mathrm{mf}} + \hat{\mathcal{D}} \left[ \sqrt{\gamma_s} \hat{p} \right] \tilde{\hat{\rho}}_{\mathrm{mf}} + \hat{\mathcal{D}} \left[ \sqrt{w} \hat{p}^{\dagger} \right] \tilde{\hat{\rho}}_{\mathrm{mf}} + \hat{\mathcal{D}} \left[ \sqrt{\gamma_p} \hat{p} \hat{p}^{\dagger} \right] \tilde{\hat{\rho}}_{\mathrm{mf}}.
\end{equation}

Our aim is to find the steady-state mean-field density matrix in the rotating frame by calculating the kernel of Eq.~\eqref{MF_MasterEq_rot}. 
We therefore need to self-consistently solve for $\omega$, $c$, and $p$ for different values of $\chi_x$.
To do this in an efficient manner, we choose a random $\omega$ (e.g., $\omega = \Gamma_c$) as well as the steady-state values from the $\chi_x = 0$ case, $c = c_0$ and $\abs{p} = \abs{p_0}$, as initial conditions of our optimizer. 
Using the values of $\omega$, $c$, and $p$ obtained from the optimizer, we can construct the mean-field master equation Eq.~\eqref{MF_MasterEq_rot} and find its kernel $\tilde{\hat{\rho}}_{\mathrm{ss}}$ in order to calculate all relevant observables. 

\begin{figure}
    \centerline{\includegraphics[width=\linewidth]{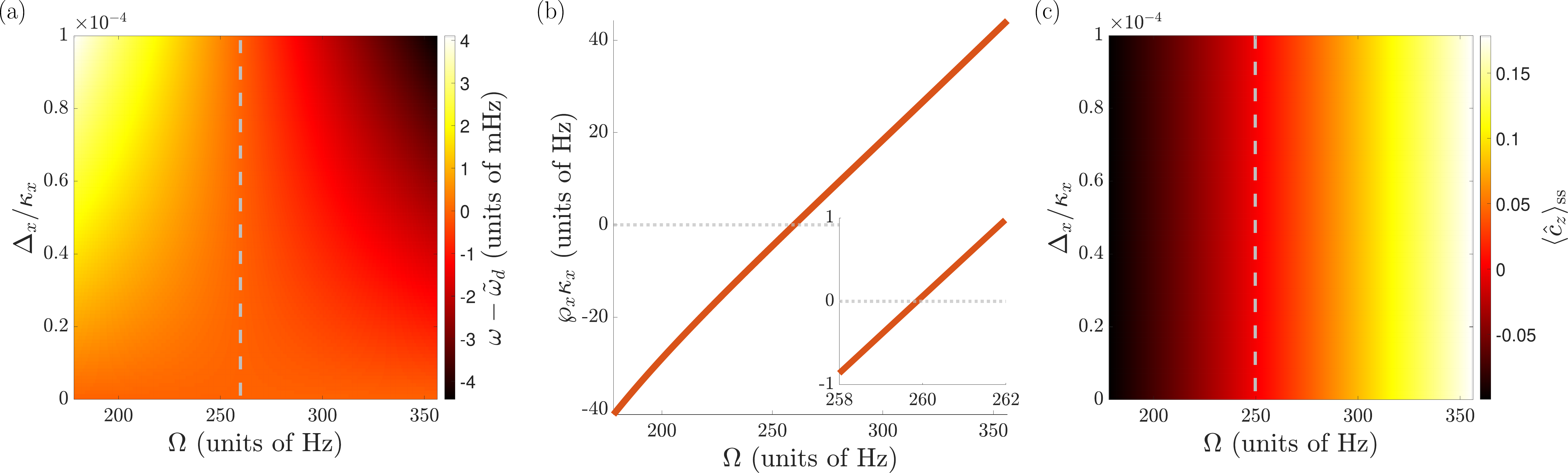}}
    \caption{(a) Laser frequency, (b) cavity pulling Eq.~\eqref{px}, and (c) inversion $\expval{\hat{c}_z}_{\mathrm{ss}}$ for the Ba parameters of Table~\ref{tb:BaParams} with $N = 10^6$.
    The inset in (b) zooms in on the $258 \, \mathrm{Hz} \leq \Omega \leq 262 \, \mathrm{Hz}$ range where the pulling crosses through zero.
    The gray dashed line in (a) and (c) show where the respective colorbars cross zero.}
    \label{MF_PullingFig}
\end{figure}
We then evaluate the pulling coefficient~\cite{Norcia3},
\begin{equation} \label{px}
    \wp_x = \frac{\omega - \tilde{\omega}_d}{\omega_x - \tilde{\omega}_d} = \frac{\tilde{\omega}_d - \omega}{\Delta_x},
\end{equation}
by varying $\chi_x$. 
In Fig.~\ref{MF_PullingFig}(a), we display the laser's detuning from the gain medium's transition frequency, $\omega - \tilde{\omega}_d$, as a function of $\Omega$ and $\chi_x \approx \Delta_x \Gamma_c / \kappa_x$ for the Ba parameters of Table~\ref{tb:BaParams} with $N = 10^6$. 
This data is then used to determine $\wp_x$ for each $\Omega$ value by taking a linear fit of vertical slices of Fig.~\ref{MF_PullingFig}(a), which gives the pulling coefficients shown in Fig.~\ref{MF_PullingFig}(b).
Throughout the entire range of Fig.~\ref{MF_PullingFig}(b), the pulling coefficient can be incredibly small when in the bad-cavity regime. 
For example, the experiment in Ref.~\cite{Norcia2} has $\kappa_x = \mathcal{O}(100 \, \mathrm{kHz})$, and so the pulling coefficient is $\abs{\wp_x} = \mathcal{O}(10^{-4})$ for the entire plot which surpasses the reported value for a continuously operated system of $0.0148$ from Ref.~\cite{Zhang4}. 
Given a vibration sensitivity of the cavity resonance of $K = \mathcal{O}(10^{-8} / \mathrm{g})$ from Ref.~\cite{Chen2}, the sensitivity to environmental noise would be reduced to~\cite{Liu} $\abs{\wp_x} K = \mathcal{O} (10^{-12} / \mathrm{g})$ which is  competitive with the most stable traditional lasers for which $K_{\mathrm{st}} = \mathcal{O}(10^{-12} / \mathrm{g})$~\cite{Robinson2,Kedar}. 

The inset of Fig.~\ref{MF_PullingFig}(b) shows that there is an entire $4$ Hz range, $258 \, \mathrm{Hz} \lesssim \Omega \lesssim 262 \, \mathrm{Hz}$, where the pulling becomes $\abs{\wp_x} = \mathcal{O}(0.1 \, \mathrm{Hz} / \kappa_x) = \mathcal{O}(10^{-6})$, giving $\abs{\wp_x} K = \mathrm{O}(10^{-14} / \mathrm{g})$, which represents a stability regime that has not been achieved for an optical laser. 
The pulling vanishes, $\wp_x = 0$, at $\Omega_0^{\wp} \approx 0.26 \, \mathrm{kHz}$; at this $\Omega$ value, we use Figs.~3(g) and~(h) of the main text (where $\Delta_x = 0$) to find a linewidth of $\Delta \nu \approx 343 \, \mu \mathrm{Hz}$ and an output power of $P_{\mathrm{ss}} \approx 3.37 \, \mathrm{pW}$. 
Lastly, we display the inversion on the lasing transition $\expval{\hat{c}_z}_{\mathrm{ss}}$ in Fig.~\ref{MF_PullingFig}(c).
Here, we find $\Omega_0^z \approx 0.25 \, \mathrm{kHz}$ (gray dashed line), exemplifying the slight shift between $\Omega_0^{\wp}$ and $\Omega_0^z$ from the final two terms in Eq.~\eqref{omegaEq}. 
Finally we note that $\Omega_0^{\wp}$ and $\Omega_0^z$ are observed to be fairly stable to small changes in $W$, $w$, and $\gamma_p$.

\bibliography{references.bib}